\documentclass[ijoc,nonblindrev]{informs3}
\OneAndAHalfSpacedXII
\usepackage{url}

\usepackage{enumerate,mathrsfs,setspace,geometry,epstopdf,bm,arydshln,amsmath,multirow,color}
\usepackage{longtable,ulem,extpfeil,extarrows,graphicx,cases,subfigure,booktabs}%
\usepackage{amssymb,lscape,diagbox}
\usepackage{graphicx}
\usepackage{verbatim}
\usepackage[colorlinks,citecolor=black]{hyperref}
\usepackage{natbib}
\usepackage{tabularx}
\usepackage{multirow}
\usepackage{multicol}
\usepackage{longtable}
\usepackage{ltxtable}
\usepackage{color}
\usepackage{rotating}
\usepackage{geometry}
\usepackage{enumerate,diagbox}
\usepackage{bm}
\usepackage{epstopdf}
\usepackage{tabu}
\usepackage{bbm}
\usepackage{booktabs}
\usepackage{setspace}
\usepackage[noend]{algpseudocode}
\usepackage[ruled,norelsize,vlined]{algorithm2e}

\usepackage{xr}

\newtheorem{defi}{Definition}
\newtheorem{prop}{Proposition}
\newtheorem{assum}{Assumption}
\newtheorem{rmk}{Remark}

\newtheorem{theo}{Theorem}
\newtheorem{coro}{Corollary}

\DeclareMathOperator*{\sta}{s.t.}

\usepackage{natbib}
\bibpunct[, ]{(}{)}{,}{a}{}{,}%
\def\bibfont{\small}%
\def\bibsep{\smallskipamount}%
\begin{document}
\RUNAUTHOR{Wang et al.} 

\RUNTITLE{The Daily Package Shipment Problem}


\TITLE{An Exact Method for the Daily Package Shipment Problem with Outsourcing}

\ARTICLEAUTHORS{%
	\AUTHOR{$\text{Zhuolin Wang}^\text{a}, \text{Rongping Zhu}^\text{a},\text{Jian-Ya Ding}^\text{b}, \text{Yu Yang}^\text{c}, \text{Keyou You}^\text{a}$}
	\AFF{$^\text{a}$Department of Automation and BNRist, Tsinghua University, Beijing 100084, China }
	\AFF{$^\text{b}$Artificial Intelligence Department, Zhejiang Cainiao Supply Chain Management Co., Ltd, Hangzhou, China, }
	\AFF{$^\text{c}$Department of Industrial and Systems Engineering, University of Florida, Gainesville, FL, 32611, U.S. }
	\AFF{\EMAIL{wangzl17@mails.tsinghua.edu.cn, zhurp19@mails.tsinghua.edu.cn, jianya.djy@cainiao.com},\\ \EMAIL{yu.yang@ise.ufl.edu}, \EMAIL{youky@tsinghua.edu.cn}}
}

\ABSTRACT{The package shipment problem requires to optimally co-design paths for both packages and  a heterogeneous fleet in a transit center network (TCN).  Instances arising from the package delivery industry in China usually involve more than ten thousand origin-destination (OD) pairs and have to be solved  daily  within an hour. Motivated by the fact that there is no interaction among different origin centers due to their competitive relationship, we propose a novel two-layer {\it localized} package shipment on a TCN (LPS-TCN) model  that exploits outsourcing for cost saving. Consequently, the original problem breaks into a set of much smaller shipment problems, each of which has hundreds of OD pairs and is subsequently modelled as a mixed integer program (MIP). Since the LPS-TCN model is proved to be \textbf{\textit{Strongly NP-hard}} and contains tens of thousands of feasible paths, an off-the-shelf MIP solver cannot produce a reliable solution in a practically acceptable amount of time. We develop a column generation based algorithm that iteratively adds ``profitable'' paths and  further enhance it by problem-specific cutting planes and variable bound tightening techniques.  Computational experiments on realistic instances from a major Chinese package express company demonstrate that the LPS-TCN model can yield solutions that bring daily economic cost reduction up to 1 million CNY for the whole TCN. In addition, our proposed algorithm solves the LPS-TCN model substantially faster than CPLEX, one of the state-of-the-art commercial MIP solvers.
}


\maketitle

%
%
%

\section{Introduction} \label{intro}
The Chinese e-commerce market has been booming, reaching  a revenue of more than 1.5 trillion  dollars in 2021\footnote{https://www.statista.com/outlook/dmo/ecommerce/china}. Consequently, more than 83.4 billion packages have been shipped in China in 2020\footnote{https://www.pitneybowes.com/us/shipping-index.html\#}, which is around 64$\%$ of the 131 billion shipped globally. Such a rapid growth not only has  posed tremendous challenge to the  package industry, but also brings numerous opportunities in cost saving and service quality improvement.

Substantial research efforts have been devoted to improving the decision making in the planning process of package express companies, which usually consists of three phases: strategic planning, tactical planning, and operational planning \citep{revelle1997design}. Strategic planning concerns long-term investment decisions  on infrastructure procurement and hub location \citep{guelat1990multimode}. Tactical planning covers mid-term decisions on the allocation of existing resources, such as service selection and traffic distribution \citep{crainic2016service}. Operational planning makes the short-term (daily) schedule for vehicles, crews, etc, to meet  fluctuating demands \citep{steadieseifi2014multimodal}. 
Generally speaking, most of the research  has focused on strategic and tactical planning, while relatively few studies consider the operational planning \citep{misni2017review}. Nonetheless, the operational planning is closely related to the actual operations performed to satisfy demands, and thus significantly impacts the transportation cost.

In this study, we consider the operational-level (daily) package shipment problem in the transit center network (TCN) commonly faced by major package express companies in China, such as Cainiao Network, SF express, etc.  Specifically, given the predicted next-day origin-destination (OD) demands,  decisions need to be made within a small amount of time (usually no more than one hour in practice),  which include the number of vehicles (of each type) to dispatch, and the corresponding routes taken to ship packages. Clearly, the routes of packages are fully coupled with those of vehicles,  which is substantially different from pure vehicle route problems \citep{toth2014vehicle}.  For the sake of cost saving, outsourcing is implemented to economically fulfill some small demands with long OD distance. Instead of outsourcing the demand of an OD pair from its origin, partial outsourcing is allowed, i.e., we first ship the demand to some assigned transit centers other than the destination and then resort to  third-party service providers to complete the shipment to the destination. Such partial outsourcing takes advantage of the shipping capacity and is expected to reduce the outsourcing cost which usually grows  proportionally w.r.t. the distance shipped. The goal is to meet all OD demands at the minimum total cost, which comprises the  transportation cost of all vehicles dispatched and the outsourcing cost.

The problem of interest involves the following two major challenges. Firstly, an existing TCN generally consists of up to one hundred centers, thousands of  arcs, and tens of thousands of OD pairs. It is extremely challenging to obtain an optimal shipment plan for the whole TCN within the acceptable time (e.g. one hour). Actually, the global optimum is impractical due to the competitive relationship among transit centers, each of which tries to maximize its own profit by delivering packages as many as possible. Secondly, the newly introduced partial outsourcing changes the traditional binary decision  of each OD demand, i.e., to outsource or not, to a much more complicated one that not only decides how to split the demand but also how to ship the split demands to the assigned transit centers. As a result, the outsourcing  decisions are highly entangled with the routing decisions, significantly complicating the problem. It is worth noticing that  there is a limit on the number of arcs for each vehicle route in practice.  This limitation is due to the fact that when vehicles arrive at a center, they have to finish unloading the packages for this center before heading to the next one. The unloading process generally takes several hours since it involves recording the origin center, updating the current center and next centers where the packages to be shipped. Hence, a path with many arcs cannot be allowed because it will significantly prolong the transportation time and  decrease the chance of on-time delivery.

To simultaneously address the two aforementioned challenges, we  first transform the global package shipment optimization to multiple local ones, which together  provides an optimal transportation plan for the whole TCN. It is worth noticing the fact that in the daily package shipment of major package express companies in China, given an OD demand, when a vehicle from the origin arrives at some transit center, loading packages originated from this transit center is usually not allowed since it involves a complex process and generally takes a long time that substantially lowers the chance of on-time delivery.   Since there is no interaction among different origin centers,  the global package shipment optimization problem over all OD pairs in the TCN is unnecessary and it suffices for each  transit center to optimize its own package shipment \textit{locally}. Thus,  we propose a localized package shipment problem on a TCN (LPS-TCN) with OD pairs corresponding to only one origin center, that is, all packages are originated from the same transit center and delivered to multiple destinations. Such a localized framework allows parallel computation for LPS-TCN problems over different origin centers and significantly reduces the difficulty of the decision making. 

Secondly, we classify the transit centers into two categories and propose a two-layer approach to model the localized package shipment optimization problem. The first layer contains one origin center and the destination centers whose packages are shipped completely by the origin itself. The second layer contains the remaining destination centers whose packages are shipped by partial outsourcing, i.e., the origin first ships these packages to some assigned transit centers in the first layer, and then resorts to third-party  to complete the remaining shipment from these assigned centers to the destinations. The novel two-layer model helps us distinguish the route for packages and vehicles. 


Finally, we obtain an LPS-TCN model on a two-layer graph, which is subsequently formulated  as a mixed integer program (MIP). Although the proposed model has significantly reduced the difficulty for the package shipment problem over the whole TCN, unfortunately, we prove that it is still strongly NP-hard to solve. Off-the-shelf solvers cannot solve the LPS-TCN  within an acceptable amount of time when there exist tens of thousands of feasible paths in the model. 
To accelerate the solution, we develop a column generation (CG) based algorithm that exploits the solution structure and further enhance it by some problem-specific cutting planes and variable bound tightening techniques.  The proposed algorithm solves the LPS-TCN substantially faster than CPLEX, one of the state-of-the-art commercial MIP solvers. More importantly,  computational experiments on realistic instances from a major Chinese package express company demonstrate that our practical model is able to reduce daily cost up to 1 million CNY  for the whole TCN.

Note that the TCN design problem at the operational level relies on full knowledge of the next-day demands, which can be predicted well by  machine learning (ML) technique \citep{ferreira2016analytics}. Hence, in this work we assume exact knowledge of the next-day demands and the problem considered is deterministic.
 
Our major contributions are summarized  as follows:
\begin{enumerate}
	\item We propose a novel LPS-TCN model over the newly constructed two-layer graph to find an optimal transportation plan at the operational level.
	\item We determine the complexity of the underlying optimization problem, i.e. finding an optimal shipment plan for vehicles and packages that jointly achieve the minimum cost.
	\item We develop a CG-based algorithm that exploits the problem structure and further enhance it by some  problem-specific cutting planes and variable bound tightening techniques.
	\item Case studies using  real-world data from a major Chinese package express company demonstrate the effectiveness of our solution approach and economic benefits of the proposed model.
\end{enumerate}

The rest of the paper is organized as follows. In Section \ref{lit_review}, we review the  related literature. In Section \ref{ProFor}, we present a detailed problem statement and the formulation of the LPS-TCN model. Section \ref{branch and price} is devoted to our proposed CG-based algorithm, where some  problem-specific cutting planes and variable bound tightening techniques are introduced. Results of the numerical study are included in Section \ref{numerical}. Finally, we conclude the paper and recommend several future directions in Section \ref{conclusion}. All proofs are provided in the online appendix.

\section{Literature Review} \label{lit_review}
The package shipment problem is becoming increasingly important and the related literature has dramatically grown during the last decade, e.g., \citet{yildiz2021hub,baloch2020strategic} for strategic planning, \citet{verma2011tactical,crainic2016service,demir2016green} for tactical planning and \citet{song2012cargo,wang2019stochastic} for operational planning. In the following, we only survey research on the package shipment problem  at the operational level, which is most relevant to our work. In particular, we focus on recent progress on models and solution methods.

\subsection{Models for the Package Shipment Problem}
The existing models for the package shipment problem can be classified as two main types: itinerary replanning (IRP) problem and fleet management (FM) problem \citep{steadieseifi2014multimodal}. The IRP problem are concerned with an optimal  response to real-time system evolution, which is not really relevant to our problem. Thus, we only focus on the FM problem.

The FM problem tackles the  movement for packages and vehicles throughout the network to ensure  on-time delivery and cost saving \citep{crainic2012fleet,chouman2015cutting}. Usually, there is a limited set of capacitated vehicles and the problem seeks an  allocation of the vehicles to demands that minimizes the transportation cost. The FM problem is one of the most fundamental problems in the TCN and has broad applications in the transportation services of road \citep{kara2004designing,osorio2015computationally}, railway \citep{yang2011railway, zhu2014scheduled}, and airline \citep{archetti2020air}.

The transit centers involved in the FM problem  usually ship packages on their own \citep{crainic2012fleet}. For an OD pair with a large demand, the origin transit center generally utilizes full truckload (FTL) to ship  packages \citep{bai2015set}, i.e.,  a dedicated vehicle is responsible for the shipment. Less-than-truckload (LTL) is also commonly used for package shipment \citep{jarrah2009large}, where packages for different destinations are consolidated and  transported by a vehicle.  LTL generally gives a higher cost saving than  FTL, as the cost per mile for the FTL is charged for the entire vehicle capacity, while LTL is computed by the actual amount of loaded packages \citep{ozkaya2010estimating}.  However, LTL takes a longer delivery time because it often requires multiple stops before the vehicle reaches the final destination \citep{xue2021hybrid}. Therefore, it is crucial to choose a proper shipment method to balance on-time delivery and cost saving.


 However,  LTL may  not  be cost-efficient enough for OD pairs with small demands but long distances. An alternative approach is to utilize outsourcing transportation to ship packages, i.e.,  transit centers outsource their package shipments to a Third-Party Logistics (3PL) company, i.e., centers hire a 3PL company to perform package shipments \citep{bardi1991transportation,aloui2021systematic}. 
 \citet{gurler2014coordinated} consider a one-warehouse $N$ retailers supply chain with stochastic demand. Inventory is managed in-house whereas transportation is outsourced to a 3PL provider. They explicitly derive the expressions of the transportation cost for the integrated joint inventory replenishment and outsourced transportation models and identify the scenarios where a 3PL transportation strategy surpasses an in-house fleet transportation strategy. \citet{cheng2014mechanism} designs a feasible plan maximizes the common profits of shippers using outsourcing service. In contrast to aforementioned works which aim to design an optimal outsourcing plan for shippers to save transportation cost,  \citet{cruijssen2010supplier} propose a new procedure that puts the initiative with the service provider,  where the logistics service provider can proactively select a group of shippers to maximize its profits.  The outsourcing transportation may increase the delivery time as the 3PL generally consolidates package shipments from different origins where multiple stops are required before vehicles reach the final destination \citep{selviaridis2008benefits,ulku2012optimal}.

In contrast to the aforementioned works which choose to ship packages totally by themselves or full outsourcing service, our proposed LPS-TCN model combines these services to seek for a transportation plan with a trade-off between cost saving and on-time delivery, i.e., given a fixed origin, packages are shipped to destinations either by the origin itself or the partial outsourcing. Different from the traditional binary decision of the OD demand, i.e., to outsource or not \citep{tang2020optimization}, the newly introduced partial outsourcing is more complicated as we need to decide not only how to split the demand but also the route of split packages. To handle the highly entangled outsourcing and routing decisions, we newly construct a two-layer graph for the LPS-TCN model to distinguish the destinations with/without partial outsourcing.

\subsection{Solution Methodology}
The FM problem is generally difficult to solve as it usually contains thousands of variables and constraints. Numerous studies investigate the methodologies to derive solutions in an acceptable computational time, e.g. \citet{barnhart1998branch,crainic2000simplex,andersen2009service,fugenschuh2015single} and \citet{jiang2017scheme,pecin2017new}.

In FM problems, arc-based models are mostly used, which are generally solved by exact algorithms such as benders decomposition approach \citep{wang2020robust,zetina2019exact}. Meanwhile, the path-based and cycle-based formulations, particularly in the TCN including thousand-level feasible paths, are also computationally interesting to study \citep{andersen2009service,jiang2017scheme}.  The cycle-based or path-based formulation outperforms the arc-based formulation in solution quality, e.g., \citet{andersen2009service} show that the cycle-based formulation exhibits gaps from 1\% to 5\% while the arc-based one yields 5\% to 20\% gap in the same solving time. However, the enumeration of all paths or cycles for large-scale network is impractical as their numbers increase exponentially with the scale. To this end, the branch and price (B\&P) method is utilized to dynamically generate the feasible paths or cycles. Moreover, violated strong linear relaxation cuts are also added in models to accelerate the algorithm \citep{alba2013branch,rothenbacher2016branch}. However, the B\&P methods are generally heavily time consuming to obtain an optimal integer solution, as designing a fast branching strategy to speed up the integer solution search is difficult \citep{alvarez2017machine}.

Due to the complexity of FM problems, heuristic and metaheuristic methods are also good choices to solve these problems. \citet{abuobidalla2019matheuristic} and \citet{sadati2021hybrid} adopt the variable neighborhood search method to find a local optimal transportation plan by exploring distant neighborhoods of a given feasible solution. Tabu Search is also a popular metaheuristic algorithm in the FM problems. \citet{xiao2018solving} set up a tabu list to allow their algorithm to accept  inferior solutions and try to avoid the local optimal solution. However, solutions of heuristic algorithms may be not reliable as there is no guarantee for the global minimum. Moreover, heuristic algorithms are unable to provide a lower bound to evaluate the quality of their solutions.

We propose a CG-based algorithm to \textit{exactly} solve our LPS-TCN model by dynamically generating feasible paths. In contrast to B\&P methods which are computational demanding to seek for optimal integer solutions due to the numerous iterations on the path generation and variable branching, the proposed algorithm takes a shorter time for optimal solutions search as it adds all columns satisfied certain conditions to the LPS-TCN model at one time. Importantly, it does not need to branch on variables and can solve the proposed model directly by MIP solvers, which also significantly reduces the computation time. Furthermore, the CG-based algorithm is further accelerated by some problem-specific cutting planes and variable bound tightening techniques.

\section{Problem Formulation} \label{ProFor}
In this section we first provide a detailed description of the LPS-TCN model, where we introduce the concept of  localization.  Subsequently,
we  describe the two-layer graph construction and present the mathematically formulation for the two-layer LPS-TCN problem.

\subsection{Problem Description}
Given a TCN,   our work seeks to answer the following  questions at the operational level

\begin{enumerate}
	\item How many vehicles of each type should be used?
	
	\item What is the best way to assign  the packages to vehicles?
	
	\item How to route the vehicles in use optimally, i.e., what are the optimal routes?
	
\end{enumerate}


A typical TCN can have more than one hundred transition centers,  thousands  of arcs, and tens of thousands of  OD pairs. Thus, finding optimal routes of the vehicles and packages over the whole TCN is complex and prohibitively time-consuming. In practice, when vehicles arrive at an intermediate transition center, loading packages originated from this center is not allowed. That is, there is no interaction among different origins, making the global optimization over all OD pairs unnecessary. Hence, it suffices to \textit{locally} optimize the package shipment for each transit center, i.e., one center is fixed as  origin,  and we find an optimal transportation plan to ship packages from this  origin to destination centers.

\hspace*{\fill} 

\noindent \textbf{Localized Model }
	\textit{We take a simple example to illustrate the localized model.} 
		
\begin{figure}[!h]
		\centering
		\includegraphics[width=0.8\linewidth]{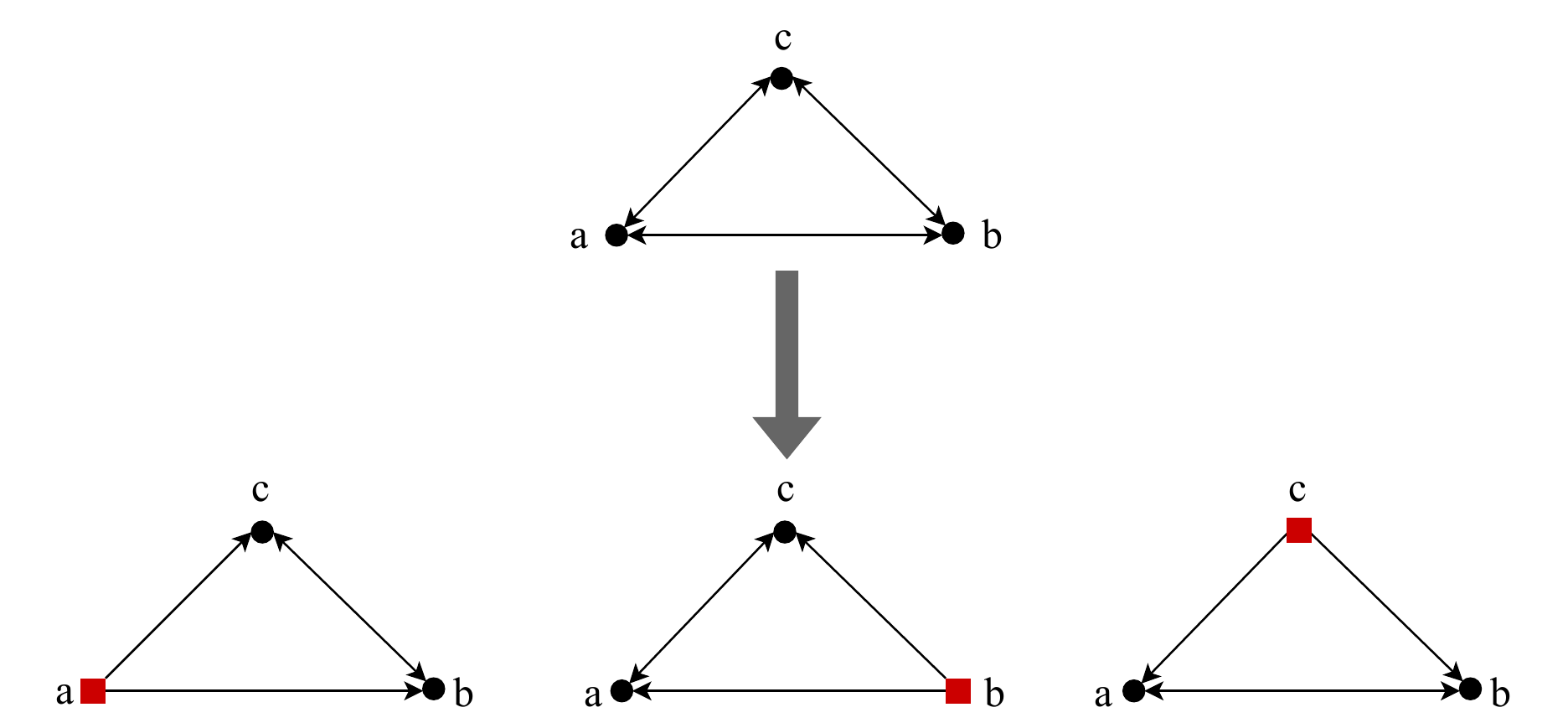}
		\caption{The localized maps for a simple TCN system.}
		\label{decen}
	\end{figure}

	\textit{Given a TCN with $3$ pairwise adjacent transit centers, where packages can be shipped between any pair in the network. The coupling of centers increases the difficulty of  global package shipment optimization in the TCN. Observe that when vehicles from one origin center arrive at an intermediate transition center, they can not load packages originated from the intermediate center, e.g., center $a$ ships packages to $b$ and $c$ via a vehicle by path $a\rightarrow b \rightarrow c$, when the vehicle arrives at $b$, it only unloads the packages for $b$ but does not load the packages needed to be shipped from $b$ to $c$. In other words, the global package shipment optimization over all OD pairs in the TCN is unnecessary since there is no interaction among different origin centers. As a result, we locally optimize the package shipment for each transit center, i.e., $a$, $b$, and $c$, which can be solved in parallel, see Figure \ref{decen} for details.}

To save transportation cost, partial outsourcing is implemented in practice to fulfill some small demands with long OD distance. To specify the destinations using partial outsourcing, we propose a two-layer graph for the localized package shipment problem with partial outsourcing, where the packages for the destination centers in first layer are shipped by the origin transit center itself, and those for destinations in the second layer are shipped by partial outsourcing.

\hspace*{\fill} 

\begin{figure}[!h]
		\centering
		\includegraphics[width=0.6\linewidth]{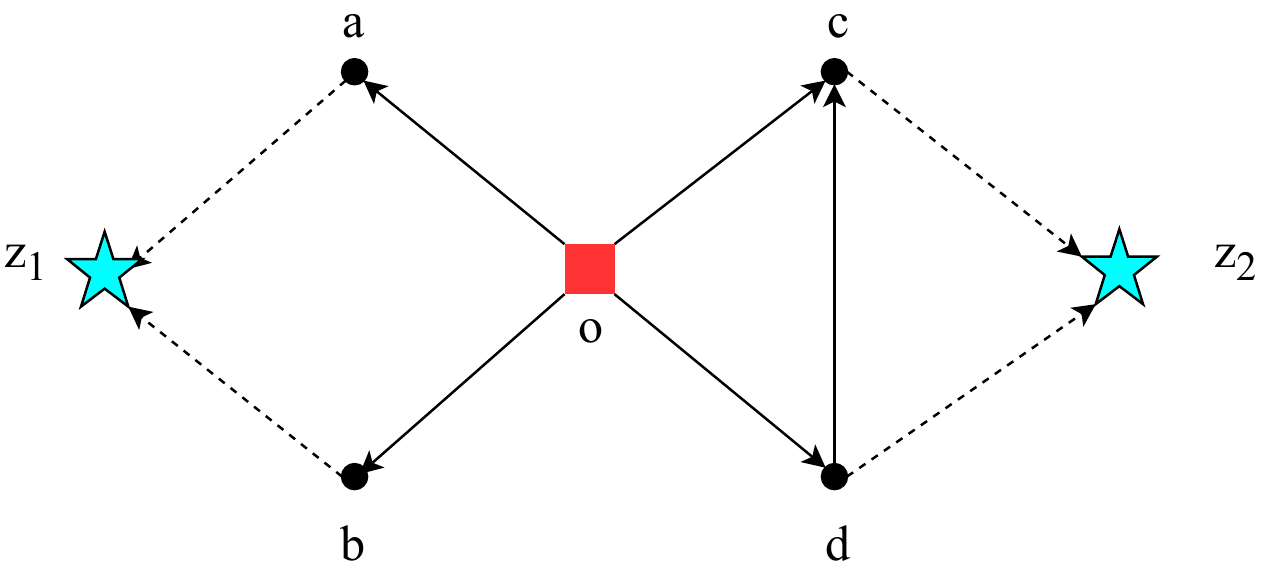}
		\caption{The Transit Center Network for packages and vehicles.}
		\label{dig1}
	\end{figure}

\noindent \textbf{Two-Layer Graph }
    \textit{We take a simple example to illustrate the constructed two-layer graph for the LPS-TCN model. Given an origin center and a set of destination centers in the LPS-TCN,  the packages must be shipped from the origin to each destination to meet its demand, see Figure \ref{dig1}, where $o$ is the origin center and others are destinations.}
	
	\textit{The origin center generally transports packages by sending vehicles on its own, e.g., $o \rightarrow d$ or $o \rightarrow c \rightarrow d$ in Figure \ref{dig1}. However, it might be uneconomical to directly ship packages to a destination with a long distance and a small amount of packages, e.g., $z_2$. In this case, the origin partially outsources the shipment to some  centers. For example,  $o$ can utilize the partial outsourcing to ship packages to transit center $z_2$ via  $c$ or $d$ or both, i.e., it first ships packages for $z_2$ to transit center $c$ or $d$ or both, and then resorts to $c$ or $d$ or both to ship these packages to $z_2$ later.}
	
	\textit{The partial outsourcing  decides how to split the  demand and how to ship the split demands to assigned transit centers, e.g., if $o$ resorts to $c$ and $d$ to ship packages to $z_2$, which paths should $o$ choose to ship the packages, the $o \rightarrow c$, $o \rightarrow d$ or $o \rightarrow c \rightarrow d$? Hence, the outsourcing and route decisions are highly entangled with each other, leading to  a complicated problem. }
		
\begin{figure}[!h]
	\centering\includegraphics[width=0.6\linewidth]{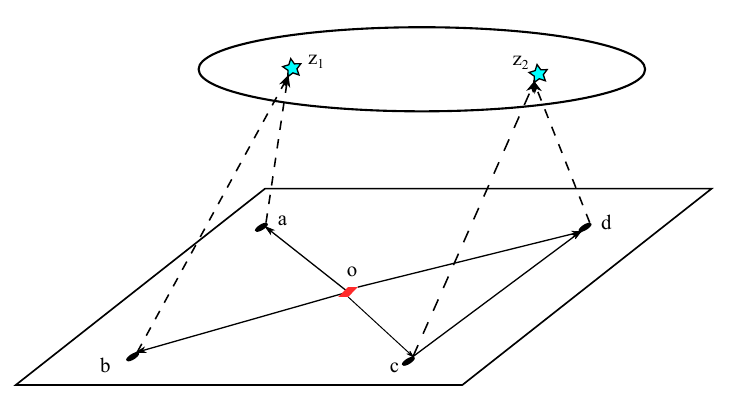}
	\caption{The two-layer graph for the TCN in Figure \ref{dig1}.}
	\label{dig2}
\end{figure}

	\textit{To deal with this challenge, we construct a novel two-layer graph (Figure \ref{dig2}) for the LPS-TCN in Figure \ref{dig1}, where one layer contains the origin and destination transit centers whose packages shipped by the origin itself, e.g, $a,b,c$ and $d$, and the other layer contains destinations whose packages are shipped by partial outsourcing, i.e., $z_1$ and $z_2$, and  dotted lines between two layers indicate that the two centers are connected, e.g. $d \dashrightarrow z_2$. }


\subsection{Notation}
We introduce the notation used throughout the paper. Let $\mathbb{R}$, $\mathbb{R}_+$, and $\mathbb{Z}_+$ denote the sets of real numbers, non-negative real numbers, and  non-negative integers, respectively. For any $I \in \mathbb{Z}_+$, we use $[I]$ to denote the set $\{1,2,\cdots, I\}$. A two-layer network $\mathcal{G}$ is represented as a directed graph $\mathcal{G} = (\mathcal{V}, \mathcal{A})$, with the origin center $o$, each node in $\mathcal{V}$ representing a destination transit center, and each arc in $\mathcal{A}$ representing a link between two centers, i.e., these centers are connected. Packages are shipped by  heterogeneous  vehicles with available vehicle types $\mathcal{K}$. For each $k \in \mathcal{K}$, let $q_k$ be the capacity of type $k$ vehicles and $c_k$ be the cost of type $k$ vehicles per kilometer. The set $\mathcal{V} = \mathcal{V}^1 \cup \mathcal{V}^2$ contains the destination centers in first layer represented as $\mathcal{V}^1$ and the ones in the second layer represented as $\mathcal{V}^2$. Similarly, the arc set $\mathcal{A} = \mathcal{A}^1 \cup \mathcal{A}^2$ also contains two parts, i.e., the set $\mathcal{A}^1 = \{(i,j)~|~ i \in \mathcal{V}^1\cup \{o\}, ~j \in \mathcal{V}^1\}$ representing  arcs in the first layer and $\mathcal{A}^2 = \{(i,j)~|~ i \in \mathcal{V}^1, ~j \in \mathcal{V}^2\}$ representing  arcs between the first and second layer. Let $\mathcal{P}$ denote the set of feasible paths in the fist layer, i.e. the path with limited number of arcs in view of the practical restriction. For each $i \in \mathcal{V}^1$, let $\mathcal{P}_i \subseteq \mathcal{P}$ be the set of paths that go through node $i$. For each $i \in \mathcal{V}$, let $\mathcal{N}_i^+=\{j \in \mathcal{V} ~|~ (i,j) \in \mathcal{A}\}$ and $\mathcal{N}_i^-=\{j \in \mathcal{V}\cup \{o\} ~|~ (j,i) \in \mathcal{A}\}$ be the sets of out-neighbors and in-neighbors of $i$,  $d_i$ be the predicted demand of  $i$, and $\mathcal{D} = \{d_i ~|~ i \in \mathcal{V}\}$ be the set of demands of all node. For each $a = (i,j) \in \mathcal{A}^1$, $\mathcal{P}_a \subseteq \mathcal{P}$ denotes the set of feasible paths containing arc $a$, $l_a$  $(l_{ij})$ is the length of arc $a=(i,j)$. And $x_a$ $(x_{ij}) \in \mathbb{R}_+$ is a continuous decision variable deciding the amount of packages on arc $a = (i,j) \in \mathcal{A}$. For each arc $(i,j) \in \mathcal{V}^2$, $c_{ij}^\prime$ is the unit outsourcing cost for shipping packages by transition center $i$  to  destination $j$. The unit is for per cubic meter (in terms of volume) of packages and per kilometer distance traveled. For each path $p \in \mathcal{P}$ and $k \in \mathcal{K}$, $y_p^k \in \mathbb{Z}_+$ is the integer variable that decides the number of  vehicles of type $k$ used on path $p$. For the sake of readability, we summarize the notation in Table \ref{Set}.

\renewcommand{\arraystretch}{0.7}
\begin{table}[!h]
\centering
	\caption{Notation}
	\begin{tabular} {c|c|l} \hline
		{Category} & {Notation} &\multicolumn{1}{c}{Description}   \\ \hline
		\multirow{13}{*}{Set} & $\mathcal{K}$    & Set of vehicle types   \\
		& $\mathcal{P}$ & Set of feasible paths to each destination $i \in \mathcal{V}^1$\\
		& $\mathcal{P}_{i}$     & Set of feasible paths containing transit center  $i \in \mathcal{V}^1$  \\
		& $\mathcal{P}_a$     & Set of feasible paths containing arc  $a$  \\
		& $\mathcal{V}^1$     & Set of destination centers  in the first layer    \\
		& $\mathcal{V}^2$   & Set of destination centers  in the second layer \\
		& $\mathcal{V}$   & $\mathcal{V} := \mathcal{V}^1 \cup \mathcal{V}^2$ \\
		& $\mathcal{A}^1$  & Set of arcs in the first layer\\
		& $\mathcal{A}^2$  & Set of arcs between the first and second layer \\
		& $\mathcal{A}$   & $\mathcal{A} :=\mathcal{A}^1\cup\mathcal{A}^2$\\
		& $\mathcal{N}^+_i$   & Set of out-neighbors of center $i$   \\
		& $\mathcal{N}^-_i$    & Set of in-neighbors of center $i$   \\
		& $\mathcal{D}$   & Set of demands\\
		\hline
			\multirow{5}{*}{Parameter}  & $l_a$ $(l_{ij})$   & Length of arc $a=(i.j)\in \mathcal{A}$     \\
		& $q_k$   & Capacity of type $k\in \mathcal{K}$ vehicle  \\
		&  $d_i$& Predicted demand of destination $i\in \mathcal{V}$  \\	
		&  $c_k$    & Unit cost of  type $k$ vehicle (per kilometer). \\ 
		& $c^\prime_{ij}$    & Unit  outsourcing cost from $j$ to $i$ (per cubic meter \\
		&  & and per kilometer) \\ 
			\hline
			\multirow{2}{*}{Decision Variable}  &  $y_{p}^{k} \in \mathbb{Z}_+$   & Number of  type $k\in \mathcal{K}$ vehicles that use path $p\in \mathcal{P}$    \\
		& $x_a$ $(x_{ij}) \in \mathbb{R}_+$ & Amount of packages transported on arc $a = (i,j) \in \mathcal{A}$  \\
		\hline		
	\end{tabular}
	\label{Set}	
\end{table}

\subsection{The Two-layer LPS-TCN Model } \label{PF-1}

Let $G = (\mathcal{V},\mathcal{A})$ be the two-layer graph with the origin node $o$, the LPS-TCN model can be formulated as the following MIP.

\begin{subequations}\label{pri-model}
	\begin{align}
	\text{min} &\sum_{a \in\mathcal{A}^1}\sum_{p\in \mathcal{P}_a}\sum_{k \in\mathcal{K}}c_kl_a \cdot y_p^k+\sum_{(i,j)\in \mathcal{A}^2}c^\prime_{ij}l_{ij}\cdot x_{ij} \label{obj} \\
	\sta & \sum_{j \in \mathcal{N}_i^+}x_{ij} - \sum_{j \in \mathcal{N}_i^-}x_{ji} =w_i, ~~\forall i \in \mathcal{V}\cup{\{o\}}, \label{demand-con}\\
	&\sum_{k \in \mathcal{K}}\sum_{p \in \mathcal{P}_a}q_ky_p^k\ge x_a, ~~\forall a \in \mathcal{A}^1,\label{capacity-con}\\
	& x_a \ge 0, ~~\forall a \in \mathcal{A},  \label{con_X}\\
	& y_p^k \in \mathbb{Z}_+, ~\forall p \in \mathcal{P}, ~~\forall k \in \mathcal{K}, \label{con_Y}
	\end{align}
\end{subequations}
where $w_i= \sum_{i \in \mathcal{V}}d_i$ if $i = o$, and  $w_i= -d_i$ if $i \in \mathcal{V}$.

Given the set $\mathcal{P}$ of feasible paths in the first layer, the proposed LPS-TCN model in \eqref{pri-model} seeks to find vehicle routes and package flows that achieve the minimum cost, which is computed in the objective function \eqref{obj} as a sum of the transportation cost in the first layer and outsourcing cost between the two layers. Constraint \eqref{demand-con} results from the  flow conservation and ensures the demand is satisfied  for each destination center, i.e., the net package flow, computed as the difference of amount of packages flowing into the center   and the amount of flowing out, equals the demand. For the origin center, it only has  outgoing flow with amount equal to the sum of all demands. Constraint \eqref{capacity-con} ensures that the total capacity of  different types of vehicles is no smaller than the amount of  package flow on each arc $a \in \mathcal{A}^1$, i.e., there is enough capacity to perform the transportation. Constraints \eqref{con_X} and \eqref{con_Y} are non-negative and integer constraints.

\subsection{The Computational Complexity of the LPS-TCN Problem} \label{set_complexity}
In this subsection, we show the computational complexity of  the LPS-TCN Problem. First, we recall the $K$-partition problem with $K\geq 3$, which has been shown to be strongly NP-complete by \cite{babel1998thek}.\\



\fbox{%
\centering
	\parbox[c]{0.95\textwidth}{%
		\begin{center}
			$K$-PARTITION Problem 
			\vspace{5 pt}
		\end{center}
		$\bm{Instance.}$ Given a list $L$ of $Km$ integers $\mu_1,\mu_2,\dots,\mu_{Km}$ with $K\geq 3$, and a bound $B \in \mathbb{Z}_+$ such that $B/(K+1) < \mu_j < B/(K-1)$ for $j = 1,\dots,Km$ and $\sum_{j=1}^{Km} \mu_j = KB$.

		\vspace{2 pt}
		$\bm{Question.}$ Can $L$ be partitioned into $m$ disjoint subsets $S_1,\dots,S_m$ such that $\sum_{j\in S_i}\mu_j = B$ for $j = 1,\dots,Km$ ?
	}%
}\\

We use this problem to analyze the computational complexity of our problem.

\begin{theo} \label{dp_np_hard}
	The problem of deciding whether there exists a feasible transportation plan for packages and vehicles where the number of arcs for each path is no larger than a given constant $n\geq 3$ and the transportation cost is no more than a given constant $C$ is Strongly NP-complete.
\end{theo}

The proof is included in Section \ref{app_proof_theo1} of the Online Appendix. The main idea is to show that  the $K$-partition problem can be polynomially reduced to our two-layer LPS-TCN problem, i.e., the  $K$-PARTITION problem can be answered by solving  an instance of the LPS-TCN model.

Theorem \ref{dp_np_hard} implies  that problem  \eqref{pri-model} is NP-hard in the strong sense. This result is not too surprising since the number of feasible paths increases exponentially as the size of the graph grows.

\section{Column Generation Based Algorithm for the LPS-TCN Model} \label{branch and price}
The LPS-TCN model is defined by the set of feasible paths $\mathcal{P}$, which is typically too time-consuming to enumerate a priori. Fortunately, if a tight enough lower bound (LB) and an upper bound (UB) of  model \eqref{pri-model} are known, many paths can be eliminated from set $\mathcal{P}$ if they do not satisfy some condition on the UB and LB  \cite[Proposition 4]{yangimproved}. In other words,  there is no need to enumerate all paths to solve \eqref{pri-model}, which inspires us  to design the following  effective CG-based algorithm. We outline the whole algorithm to solve  problem \eqref{pri-model} in Algorithm \ref{algo_all}, whose $5$ steps are detailed subsequently. By convention, the restricted master problem (RMP) in the following is referred to as the LPS-TCN model defined by the paths that has been generated so far.  

\vspace{0.1in}
\begin{algorithm}[H]
	 {\bf Input}: {A small subset of feasible paths} 
	
	 {\bf Output}: An optimal solution to the primal model \eqref{pri-model} 
	
	 {{\it Step 1. The LB Computation:} Solve the Linear programming (LP) relaxation of problem \eqref{pri-model} by CG method to get an LB and a set of paths generated in CG, denoted as $\bar{\mathcal{P}} \subseteq \mathcal{P}$  }
	
	 {{\it Step 2. The UB Computation:} Solve the RMP defined over set $\bar{\mathcal{P}}$ by an IP solver to obtain an UB  } 
	
	 {{\it Step 3. Path Enumeration:} Add all feasible paths satisfying some condition defined by the gap UB-LB to set $\bar{\mathcal{P}}$ to obtain an enlarged set $\tilde{\mathcal{P}}$    }
	
	 {{\it Step 4. Algorithm Acceleration:} Add modified rounded capacity cuts to RMP and tight the integer variable bounds  }

	 {{\it Step 5. Optimal Solution Computation:} Solve the RMP \eqref{rmp} with modified cuts and tight integer variables bounds over set $\tilde{\mathcal{P}}$ by an IP solver  }

	\caption{The algorithm for solving \eqref{pri-model}} \label{algo_all}
\end{algorithm}

Algorithm \ref{algo_all} first applies  CG method to obtain an LB by solving the LP relaxation of \eqref{pri-model} in {\it Step 1}. Then we call an IP solver to solve the RMP over the paths generated so far to obtain UB in {\it Step 2}. In the {\it Step 3},  we subsequently enumerate paths satisfying some condition defined by the gap UB - LB. To accelerate the solution, we also add some problem specific cuts and tighten the bounds of the integer variables based on a mild assumption on the network in {\it Step 4}. Finally, in {\it Step 5}, the resulting MIP is solved directly by an IP solver to obtain an optimal solution to the original problem.


\subsection{Step 1: CG Method for LB Computation}
In this subsection, we detail the CG method. It starts with a small subset of paths, which includes at least one path for each destination, such that an initial feasible solution can be obtained. Let $\bar{\mathcal{P}}_a:={\mathcal{P}}_a\cap \bar{\mathcal{P}}$, then, 
the RMP defined by $\bar{\mathcal{P}}$ is given as
\begin{subequations}
	\label{rmp}
	\begin{align}
	\text{min} &\sum_{a \in\mathcal{A}^1}\sum_{p\in \bar{ \mathcal{P}}_a}\sum_{k \in\mathcal{K}}c_kl_a\cdot y_p^k + \sum_{(i,j)\in \mathcal{A}^2}c^\prime_{ij}l_{ij}\cdot x_{ij}  \\
	\sta & \sum_{j \in \mathcal{N}_i^+}x_{ij}-\sum_{j\in\mathcal{N}_i^-}x_{ji}= w_i, ~~\forall i \in \mathcal{V}\cup{\{o\}}, \label{flow1}\\
	&\sum_{k \in \mathcal{K}}\sum_{p\in \bar{\mathcal{P}}_a}q_ky_p^k\ge x_a, ~~\forall a \in \mathcal{A}^1,\label{vehicel-n1}\\
	&x_a \ge 0, ~~ \forall a \in \mathcal{A}, \nonumber \\
	&y_p^k \in \mathbb{Z}_+, ~~\forall p \in \bar{\mathcal{P}}. \nonumber
	\end{align}
\end{subequations}
Based on an optimal dual solution of the LP relaxation of the RMP problem \eqref{rmp}, new feasible paths using different types of vehicles can be generated according to their reduced costs, which are defined as follows.

\begin{defi}
	\label{def_rc}
	For the LP relaxation of the model \eqref{rmp}, let $\pi_a$ be an optimal dual solution associated with the capacity constraint \eqref{vehicel-n1} of arc $a \in \mathcal{A}^1$, the reduced cost $r^k$ of variable $y_p^k$, which corresponds to the path $p$ using  type $k$ vehicle, is defined as
$
r^k = \sum\limits_{a\in p}(c_kl_a-q_k\pi_a).
$
\end{defi}


 After solving the linear relaxation of the RMP \eqref{rmp},  a pricing subproblem is solved to generate paths with negative reduced cost for each type of vehicles. For each type $k \in \mathcal{K}$, the subproblem can be formulated as the problem of finding the shortest path from the origin $o$ to $i \in  \mathcal{V}^1$ with arc ``distance'' $(c_kl_a-\pi_aq_k)$. Note that there is a limit on the number of arcs for a path to be feasible, the problem is actually a shortest path problem with resource constraints (SPPRC), which is known to be strongly NP-hard. The SPPRC is well studied and can be solved by dynamic programming methods \citep{bellman1958routing,ford2015flows}. We implement a dynamic programming based labeling algorithm that has been widely used in the literature  \citep{aneja1983shortest,chabrier2006vehicle,kergosien2021efficient,sadykov2021bucket}.  

Let path $p = (o,i_1,\dots, i_{n_p})$ be an elementary  path that starts from the origin, i.e., $o$, visits a set of transition centers $\mathcal{V}^p = \{i_1,\dots,i_{n_p}\}$ exactly once.  We define the label used in the labeling algorithm  as follows. 
\begin{defi}
	\label{label}
	The label $L^k_p$ associated with path $p$ using type $k$ vehicles is defined to be a $4$-tuple $L^k_p := (i_p,r^{k}_p,s_p,n_p)$, where $i_p:=i_{n_p}$ is the last transition center in path $p$,
	$r_p^{k}$ is the reduced cost for path $p$ using type $k$ vehicles, $s_p$ and $n_p$ are the length and the number of arcs of path $p$, respectively.
\end{defi}

A label $L^k_p$ is feasible if $n_p \le n$, where $n \ge 0$ is a given constant. Generating paths using type $k$ vehicles with reduced cost less than $0$  is equivalent to generating feasible labels using type $k$ vehicles with negative reduced cost, which is accomplished by the labeling algorithm. In particular, $L^k_p$ can be extend to $L^k_{p\prime} = (j,r_{p^\prime}^{k},s_{p^\prime},n_{p^\prime})$ by the following update rule, where $p^\prime = (o,i_1,\dots,i_{n_p},j)$
for $j \in \mathcal{V}^1\setminus\mathcal{V}^p$.
\begin{equation}
    \begin{aligned}
    \label{update_rule}
    & r_{p^\prime}^{k} = r^{k}_p + c_{k}l_{i_{n_p}j}-\pi_{i_{n_p}j}q_{k}\\
    & s_{p^\prime} = s_p + l_{i_{n_p}j} \\
    & n_{p^\prime} = n_p + 1.
    \end{aligned}
\end{equation}

To accelerate the labeling algorithm, we apply the dominance rule in Proposition \ref{pro:dom_rule}. 

\begin{prop}[Dominance Rule]
	\label{pro:dom_rule}
	A label $L^k_{p_1} = (i_{p_1},r^{k}_{p_1},s_{p_1},n_{p_1})$ dominates another label $L^k_{p_2} = (i_{p_2},r_{p_2}^{k},s_{p_2},n_{p_2})$, denoted by $L^k_{p_1} \prec L^k_{p_2}$, if $(\romannumeral 1)~ i_{p_1} = i_{p_2}$, $(\romannumeral 2)~ \mathcal{V}^{p_1} \subseteq \mathcal{V}^{p_2}$,  $(\romannumeral 3)~r^{k}_{p_1} \le r_{p_2}^{k}$,  $ (\romannumeral 4)~ s_{p_1}\le s_{p_2}$, and $(\romannumeral 5)~ n_{p_1} \le n_{p_2}$ hold.
	

\end{prop}



Each time a new label is obtained by the extension rule in \eqref{update_rule}, we first check its feasibility. Then,  we check whether it is dominated by other labels that have been  generated. If so, it is discarded, otherwise it is added to the label list. Lastly, we test all other labels and delete those dominated by the new one.

\subsection{Step 2: The UB Computation}
After solving the LP relaxation of model \eqref{pri-model}, we obtain an LB  and a set of feasible paths $\bar{\mathcal{P}}$. Then, we compute an UB for \eqref{pri-model} by solving the RMP problem \eqref{rmp} defined by set $\bar{\mathcal{P}}$.  Due to the moderate size of $\bar{\mathcal{P}}$, which generally consists of hundreds of feasible paths,  it can be solved fast by an off-the-shelf MIP solver, such as CPLEX.

\subsection{Step 3: Path Enumeration}

As mentioned in Section \ref{set_complexity}, there is  no need to enumerate all feasible paths. Instead, according to Proposition 4 in \citet{yangimproved} (the following Corollary \ref{reduced_cost_cut}),  it suffices to add paths using type $k$ vehicle that have reduced costs $r_p^k$ less than UB-LB into set $\bar{\mathcal{P}}$. Consequently, the RMP model \eqref{rmp} based on this enlarged $\bar{\mathcal{P}}$, denoted by  $\tilde{\mathcal{P}}$, can yield the same optimal solution as  \eqref{pri-model} with the set of all feasible paths $\mathcal{P}$. To enumerate all qualified paths,  we adopt the Yen's Algorithm \citep{yen1971finding}.

\begin{coro} \citep[Proposition 4]{yangimproved}
	\label{reduced_cost_cut}
	Given an LB and UB for the primal model \eqref{pri-model}, paths with reduced cost larger than UB-LB will not be in any optimal solution to \eqref{pri-model}, i.e., $y_p^k = 0$ if $r_p^k \ge$  UB-LB.
\end{coro}


%
%

\subsection{Step 4: Algorithm Acceleration} \label{alg_imp}

To further reduce the  computation time,  we propose to add 
some problem-specific cutting planes and tighten the  variable bounds by taking advantage of some problem structure.

\subsubsection{The Modified Rounded Capacity Cuts} \label{cut-add}
We tailor the well-known rounded capacity cuts to our problem, which ensures that the vehicles dispatched to ship packages for each destination have enough capacity to complete the shipment.

\begin{theo}
\label{valid_cut}
 The following inequality is valid for the LPS-TCN model \eqref{pri-model}
    \begin{equation}
    \label{cap-cut}
        \sum_{k \in \mathcal{K}}\sum_{p \in \mathcal{P}_i}y^k \ge \left\lceil \frac{d_i}{q^*} \right\rceil, ~~\forall i \in \mathcal{V}^1,
    \end{equation}
where $q^* =\max_{k \in \mathcal{K}} \{q_k\}$.
\end{theo}
The proof for Theorem \ref{valid_cut} is provided in Section \ref{proof:valid_cut} of the Online Appendix.

\subsubsection{Variable Bound Tightening} We utilize the problem structure to tighten the variable bounds, which is based on the following two practical assumptions.

\begin{assum}[Connectedness]
\label{assum:org}
The origin $o$ is connected to each destination center in the first layer.
\end{assum}
In practice, for each OD demand, there is always an arc linking the origin $o$ and the corresponding destination. Thus, this assumption is  always satisfied.

\begin{assum}[Triangle Inequality]
\label{assum-tra}
For any arc $(i,j)\in \mathcal{A}^1$,  $l_{ij} \leq  l_{ik}+l_{kj}$  holds for  $\forall (i,k),   (k,j) \in \mathcal{A}^1$. 
\end{assum}

In practice this assumption may not always be satisfied. But it is a common assumption in the context of vehicle routing and matches reality in most cases.

Let $p =(o,i_1,\dots,i_{n_p})$ be a path and $\{\bar{z}^{ki_j}_p\}_{j=1}^{n_p}$ be the amount of packages delivered to node $i_j$ via path $p$ by type  $k$ vehicle. Under Assumptions \ref{assum:org} and \ref{assum-tra}, we can tighten the variable bound by the following Theorem \ref{var_red}. 
\begin{theo}
\label{var_red}
    Under Assumption \ref{assum:org} and \ref{assum-tra}, for any optimal solution $(\bar x, \bar y)$ to \eqref{pri-model}, we have $\bar y^k\leq 1$ for each $p$ with $n_p \ge 2$. Furthermore,  $\bar{z}^{ki_j}_p \leq  q_k$ for each node $i_j$ in path $p$.
\end{theo}

Theorem \ref{var_red} is proved by contradiction in Section \ref{proof:var_red} of the Online Appendix. The main idea is to show that if the number of type $k$ vehicles on path $p$ with $n_p \ge 2$ is larger than $1$, i.e., $\bar{y}_p^k > 1$,  we can always request one of the vehicles to travel a different path $p^\prime$. The result will not increase the cost and will still satisfy all the demands. More precisely,  we have $\mathcal{V}^{p^\prime} \subset \mathcal{V}^p$ and $l_{p^\prime} \leq  l_p$.

%
Without loss of generality, we assume that the capacity of vehicles $\{q_k\}_{k \in \mathcal{K}}$ are in an ascending order, that is, $q_i < q_j, \forall ~i \leq  j, ~i,j \in \mathcal{K}$. Under this assumption, except for the vehicle with the maximal capacity, i.e. $k = |\mathcal{K}|$, we can find an upper bound for the number of type $k$ vehicle on path $p$ with only one arc by solving an integer program.

\begin{theo}
\label{bound:one_arc}
We can find an optimal solution $(\bar x,\bar y)$ to problem \eqref{pri-model} such that $\bar{y}_p^k$ is no larger than the optimal value of the following integer program for each $k \in \mathcal{K}\setminus\{|\mathcal{K}|\}$ and $p$ with $n_p = 1$.
\begin{subequations}
\label{add_bound}
    \begin{align}
        {\rm min.} &~~~~ u_k \nonumber \\
        {\rm s.t.} & ~ \sum_{i \in \mathcal{K}, i > k}c_iv_i \le c_ku_k \label{con:cost} \\
        &~ \sum_{i \in \mathcal{K},i > k}q_iv_i \ge q_k u_k \label{con:cap}\\
        &~ u_k \in \mathbb{Z}_+,~ v_i \in \mathbb{Z}_+, \forall i \in \mathcal{K}, i > k, \nonumber
    \end{align}
\end{subequations}
where $u_k$ is an integer variable counting the number of type $k$ vehicle on path $p$, $v_i$ is the integer variable counting the number of type $i \in \mathcal{K}, i > k$ vehicle on path $p$. The constraint  \eqref{con:cost} requires to find a group of vehicles with capacity larger than $q_k$ such that their total cost is no larger than $c_ku_k$. Constraint \eqref{con:cap} requires that the total capacity of this group is no less than $q_ku_k$.
\end{theo}

Theorem \ref{bound:one_arc} is proved in Section \ref{proof:bound:one_arc} of the Online Appendix. Note that problem $\eqref{add_bound}$ depends only on the vehicle type, and thus the number of type $k$ vehicles on different path $p$ has the same upper bound.  Hence, we only need to solve $|\mathcal{K}|$ problems to obtain all  upper bounds for $y$ variables. Generally, the number of the vehicle types is small and  the problem \eqref{add_bound} is easily solved by an IP solver.

\subsection{Step 5: Optimal Solution Computation}
We have enumerated all necessary paths (columns),  added problem specific cuts to the proposed model, and tightened the bounds of integer variables. The resulting MIP can be solved directly to obtain an optimal solution to the original problem. In this step, we adopt one of the state-of-the-art MIP solvers  CPLEX to compute an optimal solution to problem \eqref{pri-model}.

\section{Numerical Results}\label{numerical}
In this section, we illustrate the superiority of the CG-based Algorithm \ref{algo_all} compared with CPLEX's default method and test the performance of the proposed LPS-TCN model using  real-world instances. The code is implemented in C++  and all experiments are performed on a workstation running  Ubuntu 20.04 with AMD Ryzen Threadripper 3990X 64-core, 128-thread processor and 128 GB RAM.  CPLEX 12.8  is employed to solve the involved MIP and LP models. {We limit the number of arcs on feasible paths to $3$, which is consistent with the path limitation in a major Chinese package express company.}

\renewcommand{\arraystretch}{1.2}
\begin{table}[!h]
	\caption{Selected origin transit centers}
	\centering
	\begin{tabular} {c|l} \hline
		Scale & \multicolumn{1}{c}{City} \\ \hline
		Small & Fuyang, Zigong, Tianjin, Luoyang, Taizhou,
			   Wuhu, Bengbu, Changchun, \\
			   & Liuzhou, Guiyang \\
		Middle& Fuzhou, Shengyang, Linhai, Haerbin, Kunming\\
		Large& Wenzhou, Shanghai, Wuhan, Beijing, Zengcheng \\
		\hline		
	\end{tabular}
	\label{city_set}	
\end{table}

The raw data used in our experiments comes from a major Chinese package express company and has been desensitized for data security. In total, there are  $20$ instances generated by the following steps. $1)$ Select $20$ transit centers as origins, which contain $10$ small-scale, $5$ middle-scale, and $5$ large-scale centers; $2)$ Build the corresponding two-layer graph for the LPS-TCN based on the ``Current-Next-Final" (CNF) network provided by the company, where  $C$ is one of the selected origins, $N$ is the set of next centers to which the packages are shipped, $F$ is the set of final center for packages. We set the $N$ centers as destination nodes in the first layer and $F$ centers as destination nodes in the second layer. Moreover, the origin is connected with each destination centers in the first layer and if the distance between two destination centers in the first layer is less than a given constant, e.g., $500$km, these centers are connected by arcs. Then, we obtain a two-layer graph for the proposed LPS-TCN model. $3)$ We collect package flows on the CNF network for each of the $20$ origins in the LPS-TCN over a week spanning from September 13, 2021 to September 19, 2021. The OD demands are calculated by  package flows. Consequently, we have $20*7 = 140$ different two-layer graphs for our LPS-TCN models.

\begin{rmk}
Figure \ref{fig:cndmap} is an example that illustrates step $2$, where the red parallelogram is current center, triangles are  next centers and hexagons are final centers. We transform the CNF network to a two-layer graph, i.e., set the triangles as  destination centers in the first layer and hexagons as centers in the second layer. Moreover, as the distance between $N_2$ and $N_3$ is less than $500$km, they are connected in the two-layer graph. 
	
	\begin{figure}[!ht]
		\centering
		\includegraphics[width=0.8\linewidth]{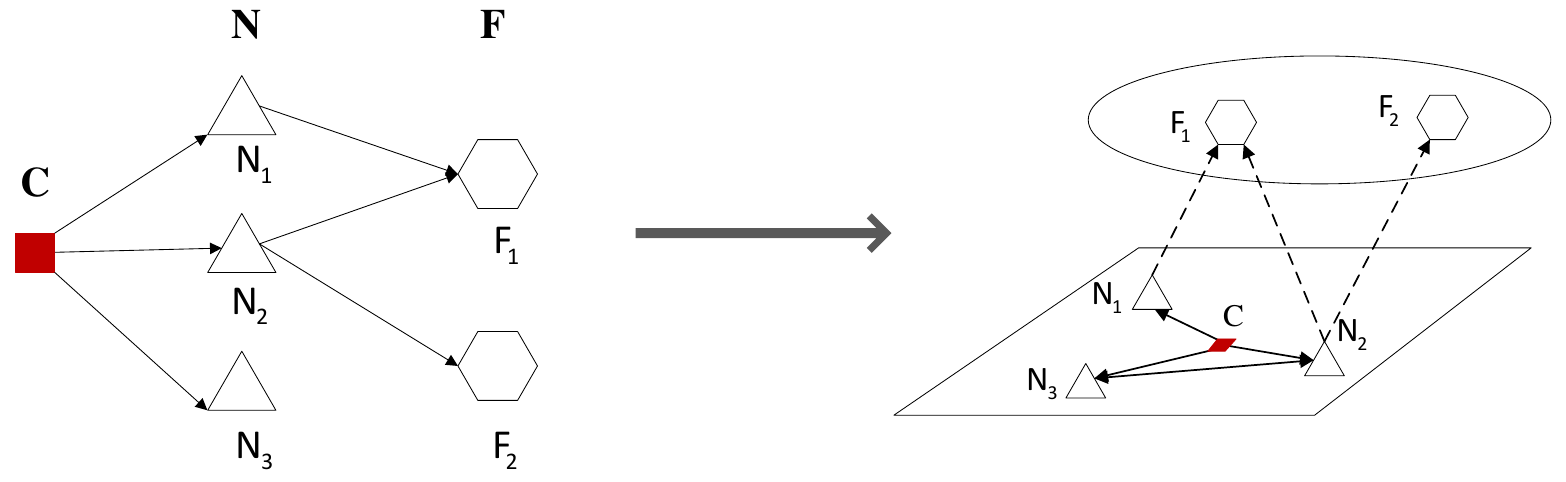}
		\caption{The CNF network and the corresponding two-layer graph for the LPS-TCN model.}
		\label{fig:cndmap}
	\end{figure}
		
\end{rmk}


There are 4 vehicle types, i.e., $\mathcal{K} = \{1, 2, 3, 4\}$,  in our experiments, and the corresponding parameters used are summarized in Table \ref{para_set}. The distance between transit centers are captured from the AMAP which provides a live traffic data interface\footnote{https://restapi.amap.com/v3/direction/driving}. 

\renewcommand{\arraystretch}{1.2} 
\begin{table}[!h]
\centering
	\caption{Parameters Setting}
	\begin{tabular} {c|l} \hline
		Parameter & \multicolumn{1}{c}{Setting}  \\ \hline
		$q_k $     &$q_1 = 65, q_2 = 90, q_3 =  130, q_4 = 175$, with unit $\text{m}^3$ \\
		$c_k$&  $c_1 = 4.1, c_2 = 4.7, c_3 = 6.5, c_4 = 7.5$, with unit CNY/km\\
		$c^\prime_{ij}$& $0.06$ with unit $\text{CNY}/(\text{km} \cdot \text{m}^3)$ \\
		\hline		
	\end{tabular}
	\label{para_set}	
\end{table}

\subsection{The Optimal Vehicle Routes in Different Transit Centers}

\begin{figure}[h] \centering
	\subfigure[Illustration of vehicle routes for Luoyang transit center.]{ \label{fig:luoyang}
		\includegraphics[width=0.45\linewidth]{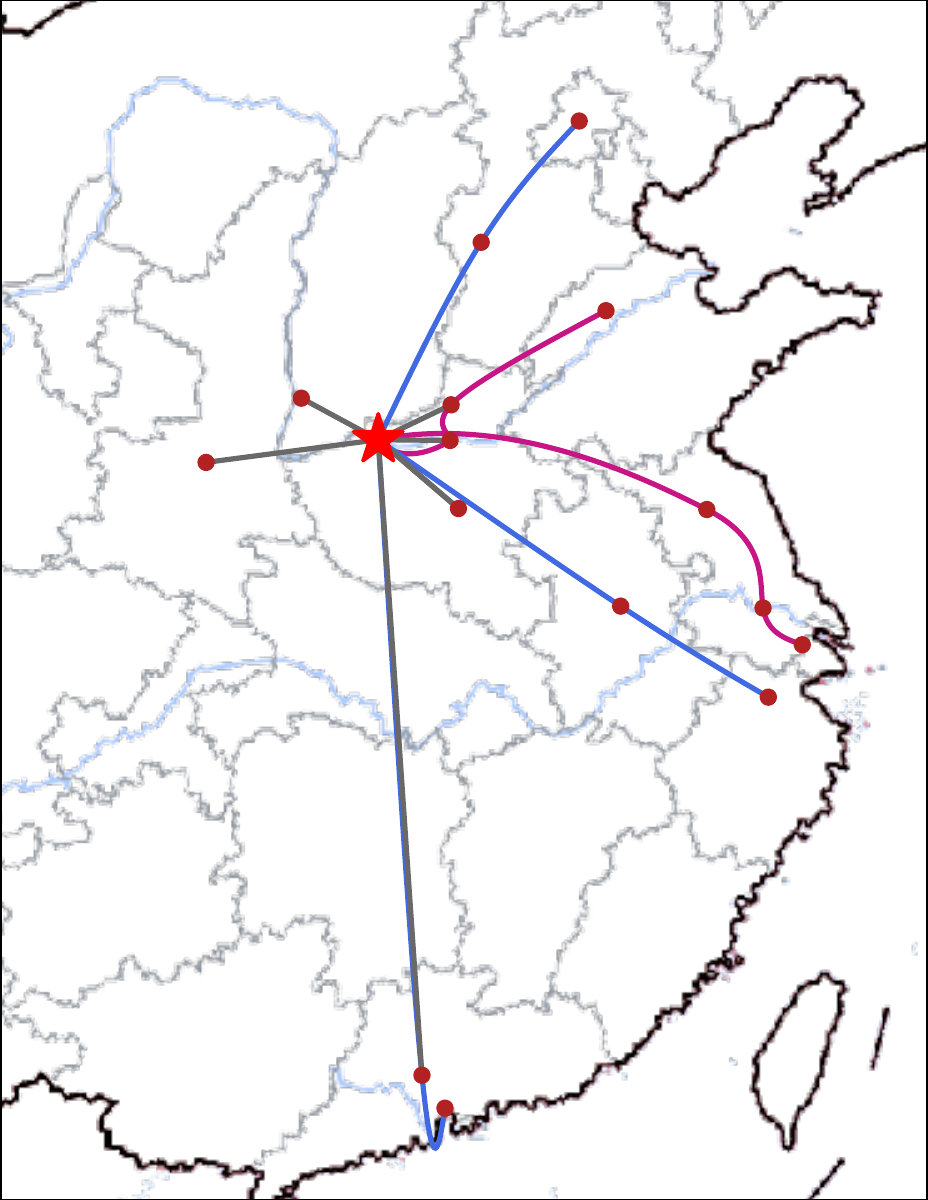} 
	}
	\subfigure[Illustration of vehicle routes for Wuhan transit center.]{ \label{wuhan}
		\includegraphics[width=0.48\linewidth]{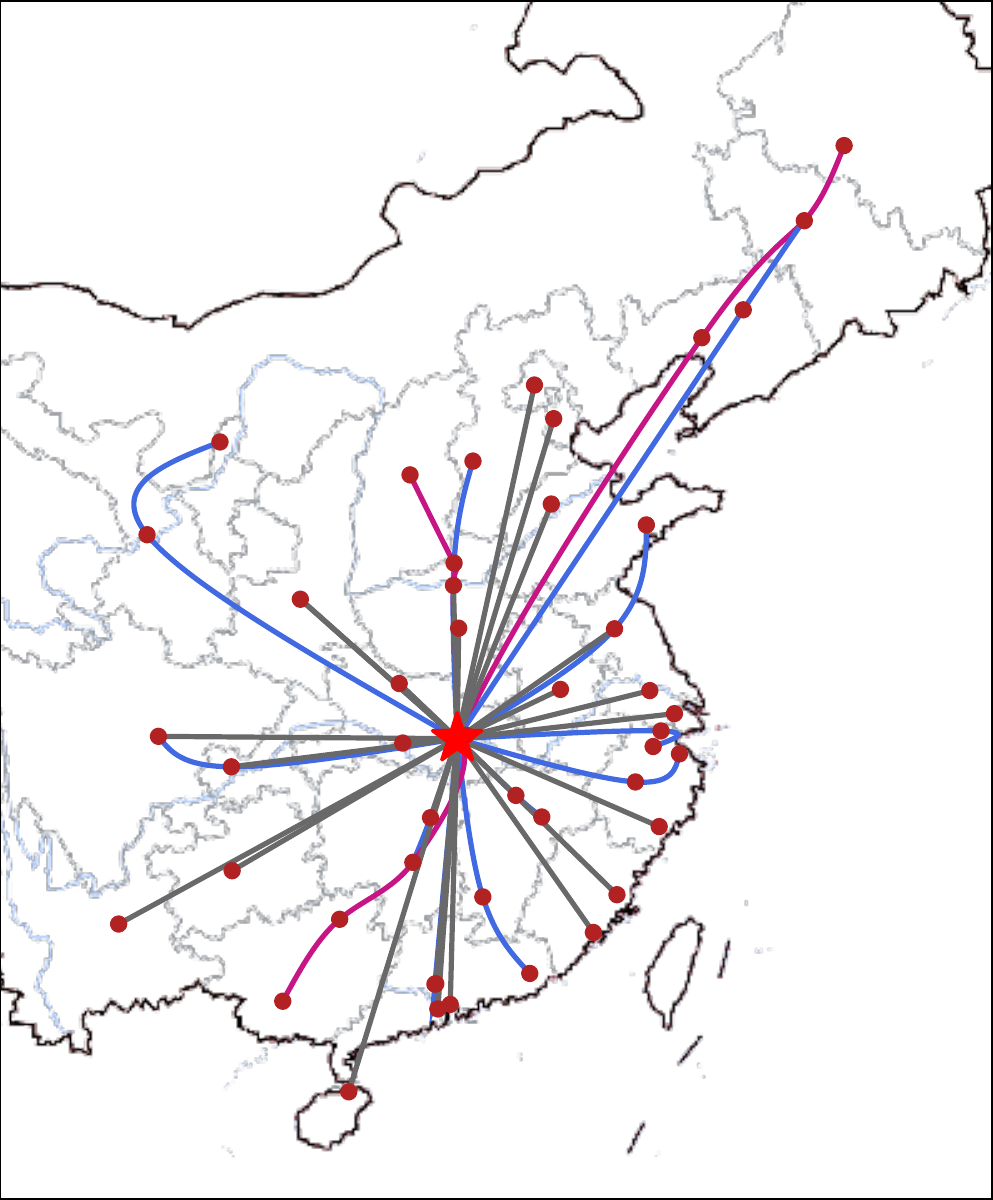} %
	}
	\caption{The vehicle routes in different transit centers}
	\label{fig:route}
\end{figure}

Figure \ref{fig:route} shows the vehicle route decisions of the proposed LPS-TCN model for two different origins, where ``Luoyang" is a small-scale origin transit center and ``Wuhan" is a large-scale one. The red stars and dots represent the origin and destination centers, respectively. The gray, blue, and purple lines indicate that the paths have one, two, and three arcs respectively. 

We can observe from Figure \ref{fig:route} that the number of  paths with one arc (gray lines) is the largest, which indicates that most vehicles are responsible for single shipments, i.e., the vehicle only ships the packages to one destination. Moreover, the optimal routes  obtained by our model match the true situation. 

\subsection{Performance of the Proposed LPS-TCN Model} \label{sec:cost_com}
Numerical experiments on the aforementioned $20$ instances are conducted to evaluate the performance of the proposed LPS-TCN model. We first compare the averaged transportation costs over $7$ days of the LPS-TCN model  with the averaged real costs. Then, we compare the cost of the two-layer graph LPS-TCN model with that of a transportation plan which also takes advantage of partial outsourcing. The purpose of the second set of experiments is to verify that in addition to partial outsourcing, the optimization of  package flows and vehicle routes also help to drive the transportation cost down.

\begin{rmk}
	 Although the collected data does not provide the real transportation cost directly because it does not include  vehicle routes, we can derive them by solving an integer program  based on the given package flows. Consequently, we  obtain the true transportation cost. 
\end{rmk}

\begin{rmk}
The additional transportation plan with outsourcing is derived as follows. We start from the LPS-TCN model over the two-layer graph constructed from the CNF network, and fix its package flows according to the real plan. Then, we derive vehicle routes in the first layer based on the fixed package flows and assume that the packages to centers in the second layer are handled by partial outsourcing. Thus, the transportation cost can be computed accordingly.
\end{rmk}

\begin{figure}[!htb] \centering
	\subfigure{ \label{fig:large_cost1}
			\includegraphics[scale = 0.45]{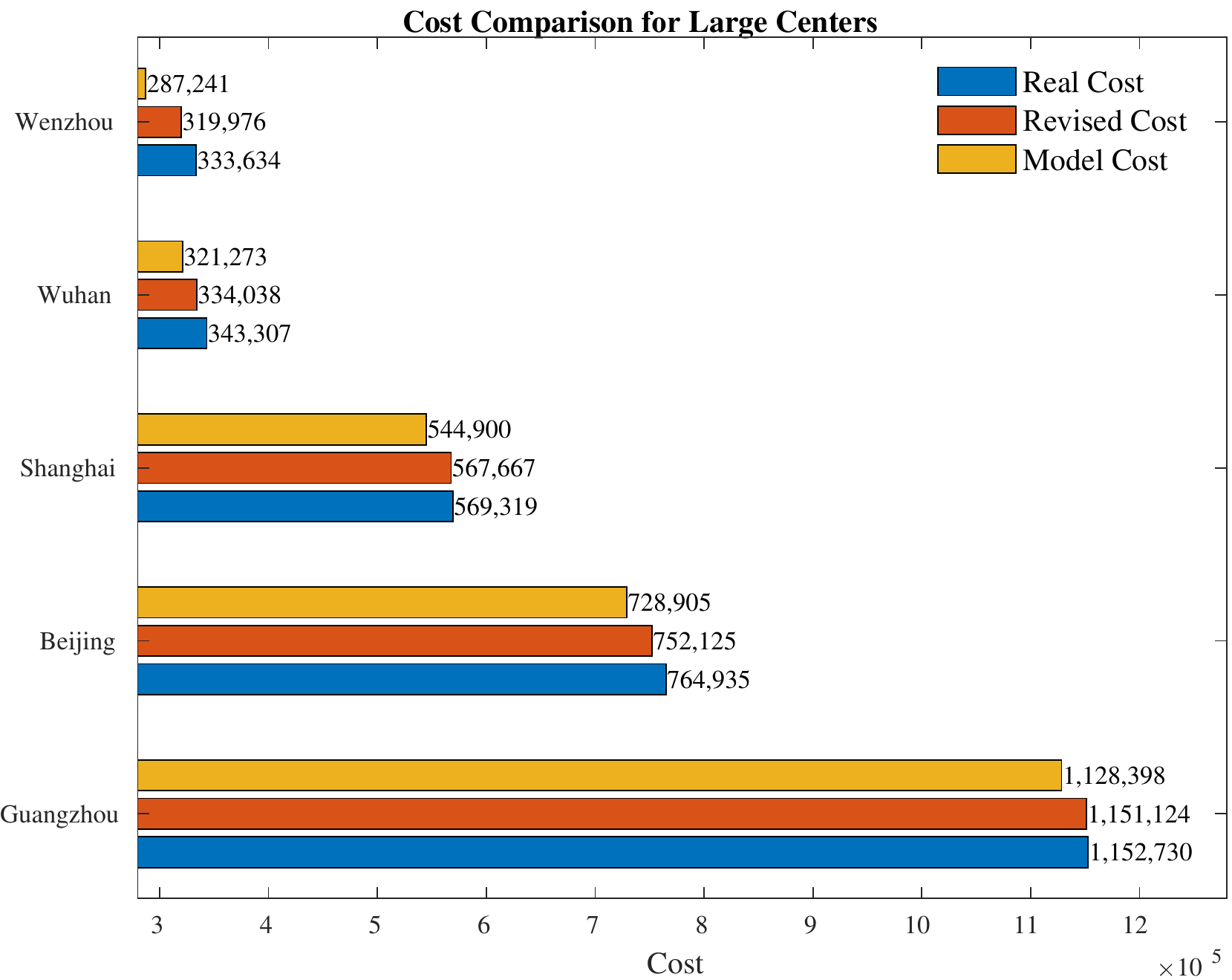} %
	}
	\subfigure{ \label{fig:mid_cost1}
		\includegraphics[scale = 0.45]{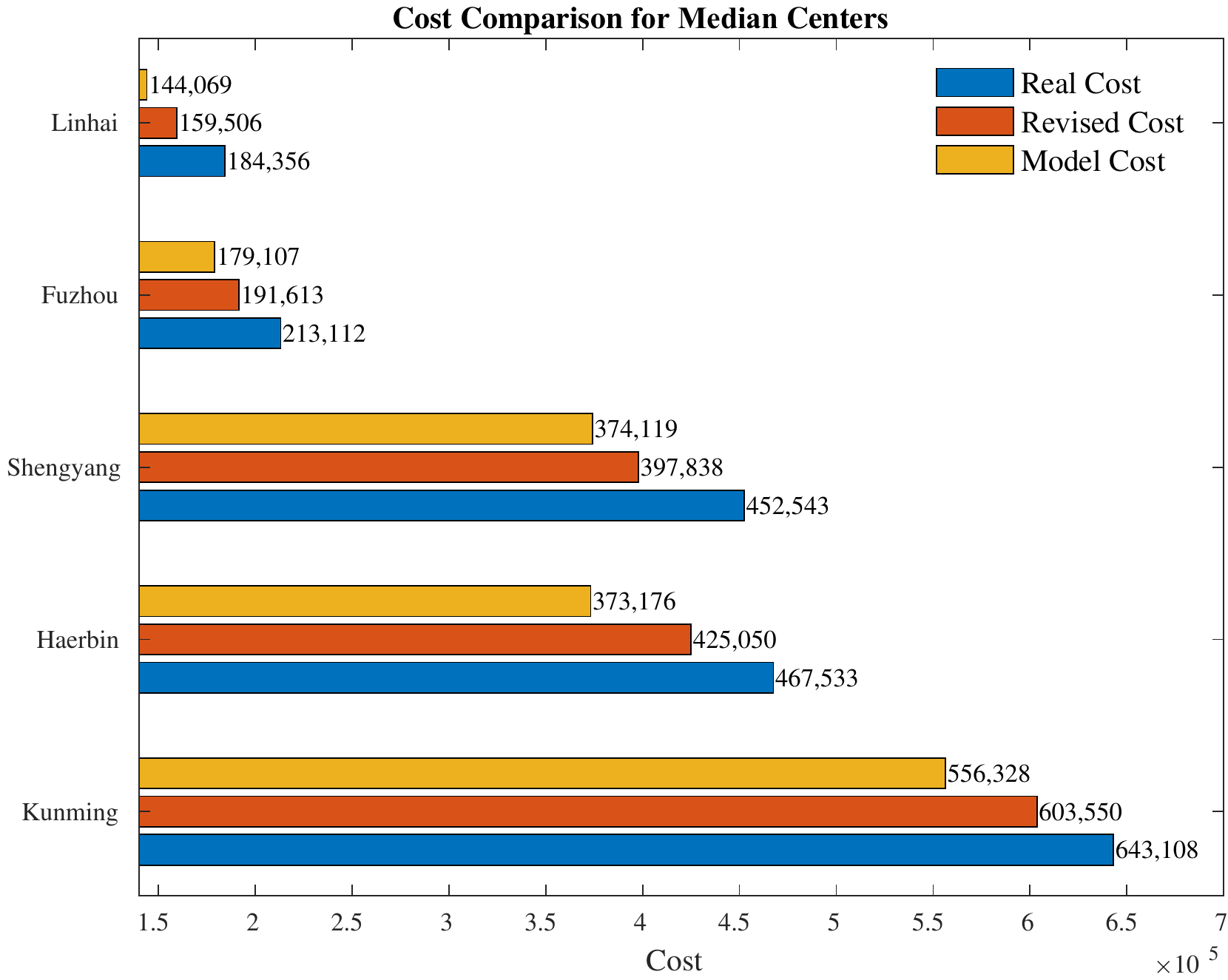} %
	}
	\subfigure{ \label{fig:small_cost1}
		\includegraphics[scale = 0.45]{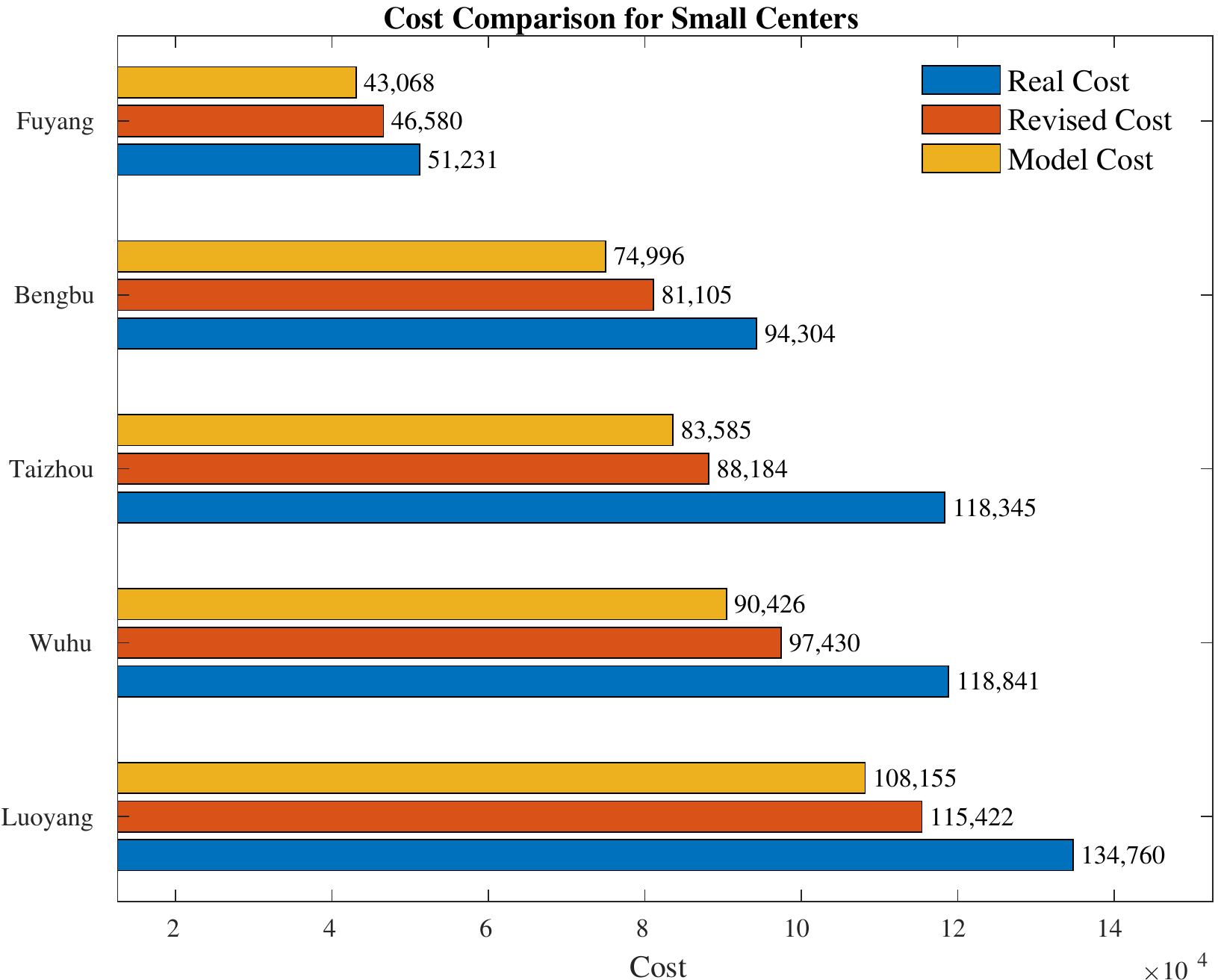} 
	}
	\subfigure{ \label{fig:small_cost2}
		\includegraphics[scale = 0.45]{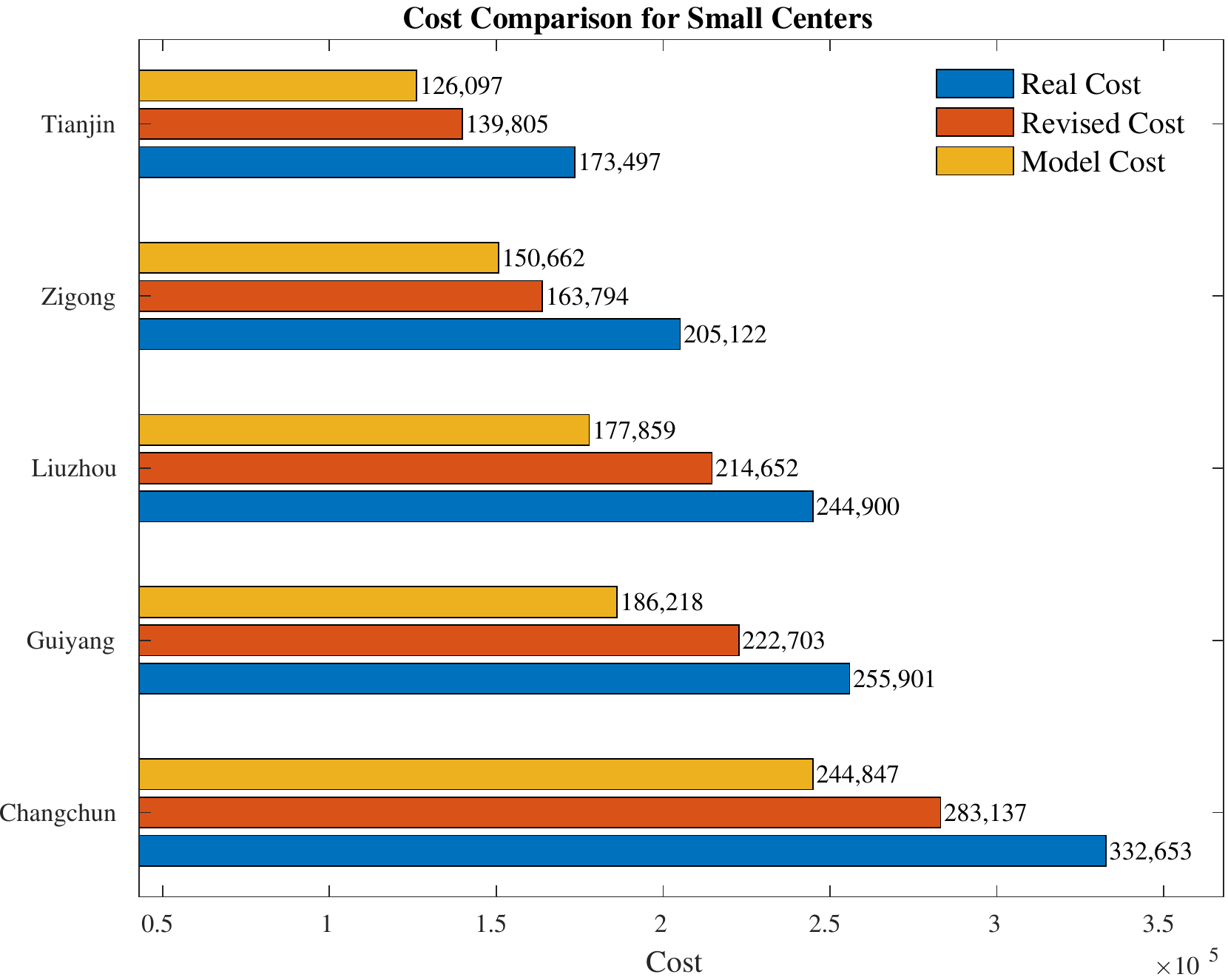} 
	}
	\caption{Averaged cost (CNY) of different methods}
	\label{fig:cost_com1}
\end{figure}

Results are reported in Figure \ref{fig:cost_com1}, where ``Real Cost" represents the true transportation cost, ``Revised Cost" is the cost of the derived transportation plan based on fixed package flows, and ``Model Cost" is the  objective value of the best solution  to the LPS-TCN model at termination.


From Figure \ref{fig:cost_com1},  one can observe that our LPS-TCN model yields solutions that bring significant  economic benefits, i.e., a daily cost reduction up to tens of thousands of CNY for each transit center. Consequently, the daily cost reduction for the whole TCN with hundreds of transit centers is more than \textbf{one million}  CNY. An interesting observation is that the cost reduction depends on the scale of  transit center and the larger the center, the smaller the cost reduction. The intuition behind is that large-scale transit centers  generally have large OD demands and tend to ship most of the packages by  themselves. Hence, the packages to be partially outsourced is smaller,  which leads to a smaller cost reduction. Figure \ref{fig:cost_com1} also shows that the LPS-TCN model exhibits better performance in terms of the transportation cost than the derived transportation plan with outsourcing, which indicates that our proposed model can additionally provide better package and vehicle routes. In summary, our LPS-TCN model can reduce the costs not only by providing better routes for packages and vehicles but also by taking advantage of partial outsourcing.

\subsection{Performance Analysis of the CG-based Algorithm}\label{subsection:cplex}
The goal of this subsection is to perform a comprehensive comparison of our CG-based algorithm and the CPLEX's default method. Experiments  on real instances for different origins with scale parameters ($\mathcal{V}^1$, $\mathcal{V}^2$) over seven days are conducted. The $(D^1,D^2)$ denotes the total demands in the first and second layer respectively. The computation is  terminated once the CPU time reaches 1,800 seconds (half an hour).

Table \ref{tab:cplex} reports the average computational time (Time), the optimality gap at termination (Gap), and the percentage of instances solved to optimality within the time limit (Opt), where the first two metrics are averaged over seven instances. Columns ``CPLEX" and ``CG" represent the performance of CPLEX's default method and the proposed CG-based method respectively.

\renewcommand{\arraystretch}{0.65}
		\begin{longtable}{ccccccccccc}
			\caption{Performance analysis of the CG-based algorithm} 	\cr
			\hline
			\multirow{2}[0]{*}{Origins} & \multirow{2}[0]{*}{$\mathcal{V}^1$} & \multirow{2}[0]{*}{$\mathcal{V}^2$} & \multirow{2}[0]{*}{$D^1$} & \multirow{2}[0]{*}{$D^2$} & \multicolumn{2}{c}{Time} & \multicolumn{2}{c}{Gap} & \multicolumn{2}{c}{Opt} \cr
			\cmidrule(r){6-7}\cmidrule(r){8-9}\cmidrule(r){10-11}
			& & &  &  & CPLEX  & CG & CPLEX  & CG & CPLEX  & CG \cr \midrule
			{Fuyang}  & 7  & 103 & 333 & 595 & 0.77 & 0.57 & 0  & 0  & {100\%} & {100\%}\cr
			\midrule
			{Zigong}  & 11  & 108 & 622 & 1,217 & 0.30 & 0.24 & 0  & 0  & {100\%} & {100\%}\cr
			 \midrule		 
			{Tianjin}  & 15  & 104 & 721 & 1,232 & 1.95 & 1.30 & 0  & 0  & {100\%} & {100\%}\cr
		   \midrule	
			{Luoyang}  & 17  & 104 & 737 & 1,178 & 12.78 & 8.62 & 0  & 0  & {100\%} & {100\%}\cr
		   \midrule							
			{Taizhou}  & 15  & 102 & 970 & 1,003 & 62.90 & 56.08 & 0  & 0  & {100\%} & {100\%}
			 \cr \midrule
			{Fuzhou} & 19	& 100	& 1,675	& 1,585	& 21.76	& 14.34 & 0  & 0  & {100\%} & {100\%} \cr
			 \midrule
			{Bengbu}& 19	& 98	& 1,255	& 748	& 30.14	& 22.69 & 0  & 0  & {100\%} & {100\%}\cr
			\midrule	
			{Changchun}  & 23  & 100 & 1,192 & 1,246 & 36.86 & 26.09 & 0  & 0  & {100\%} & {100\%}\cr
			\midrule
			{Shenyang} & 23	& 100	& 2,112 & 2,324 & 24.21 & 15.97 & 0  & 0  & {100\%} & {100\%} \cr		\midrule
			{Guiyang} & 25	& 90 & 1,177 & 1,297 & 52.07 & 29.14 & 0  & 0  & {100\%} & {100\%}\cr \midrule
			{Wuhu} & 22	& 92	& 1,682 & 868 & 291.15 & 147.21 & 0  & 0  & {100\%} & {100\%}\cr\midrule
			{Liuzhou}  & 29  & 91 & 2,191 & 1,047 & 100.77 & 40.42 & 0  & 0  & {100\%} & {100\%}\cr \midrule
			{Linhai} & 30	& 92	& 1,436	& 910 & 116.53 & 88.78 & 0  & 0  & {100\%} & {100\%}\cr
			\midrule
			{Haerbin} & 28	& 96	& 3,075	& 1,569	& 117.06 & 75.16 & 0  & 0  & {100\%} & {100\%}\cr \midrule
			{Kunming} & 33	& 86	& 3,753	& 2,645	& 1,583.68	& 1,200.48 & 0.33\%  & 0.21\%  & {14\%} & {42\%}\cr \midrule	
			{Wenzhou} & 52	& 73	& 3,859	& 1,121	& 1,357.16	& 650.56 & 0.04\%  & 0.01\%  & {42\%} & {86\%}\cr \midrule 			
			{Shanghai} & 66	& 58	& 12,550	& 1,532	& 1,800	& 1,575.64 & 0.25\%  & 0.09\%  & {0} & {14\%}\cr \midrule		
			{Wuhan} & 48	& 76	& 7,707	& 1,736	& 1,800	& 1,800 & 0.67\%  & 0.50\%  & {0} & {0}\cr \midrule						
			{Beijing} & 66	& 59	& 13,517	& 1,808	& 1,800 & 1,800 & 0.49\%  & 0.29\%  & {0} & {0}\cr \midrule	
			{Zengcheng} & 69	& 57	& 13,518	& 1,621	& 1,800	& 1,800 & 0.39\%  & 0.32\%  & {0} & {0}\cr		
			\bottomrule
			\label{tab:cplex}
		\end{longtable}

As shown in Table \ref{tab:cplex}, the average computational time of the CG-based Algorithm \ref{algo_all} is generally much smaller than that of CPLEX. In particular, for small and middle-scaled instances,  the computational time of Algorithm \ref{algo_all} is only 50\% of that of  CPLEX. For those instances that cannot be solved within the time limit,  Algorithm \ref{algo_all} achieves a significantly smaller optimality gap. For example, when the origin is Wuhan, although both algorithms fail to obtain an optimal solution, the Gap of Algorithm \ref{algo_all} is smaller than that of CPLEX; when the origin is Wenzhou, CPLEX only solves three instances to optimality, while Algorithm \ref{algo_all} proves optimality for six of the seven instances.

\subsection{Sensitivity Analysis} \label{sec:sensitivity}

As mentioned in Section \ref{intro}, we should place centers with small demands and meanwhile large OD distances in the second layer to facilitate cost saving. The corresponding thresholds for demands and OD distances can affect the solution quality. 
In this subsection, we test the impact of such thresholds (decided by the ratio $0 \le \theta \le 1$) on our two-layer LPS-TCN model. Details of the experiments are summarized as follows. We first select three  small-scale origin centers, three middle-scale ones, and another three large-scale ones, as test instances. For each origin, we sort its OD demands in an ascending order and OD distances in a descending order. Then, the destination centers with OD demands and distances ranking within the top $\theta$ are selected as nodes in the second layers and the others are placed  in the first layers. The threshold $\theta$ tested are from $\{0.9,0.8,0.7,0.6,0.5,0.4\}$.

\begin{figure}[h] 
\centering
	\subfigure{ \label{fig:small}
		\includegraphics[width=0.30\linewidth]{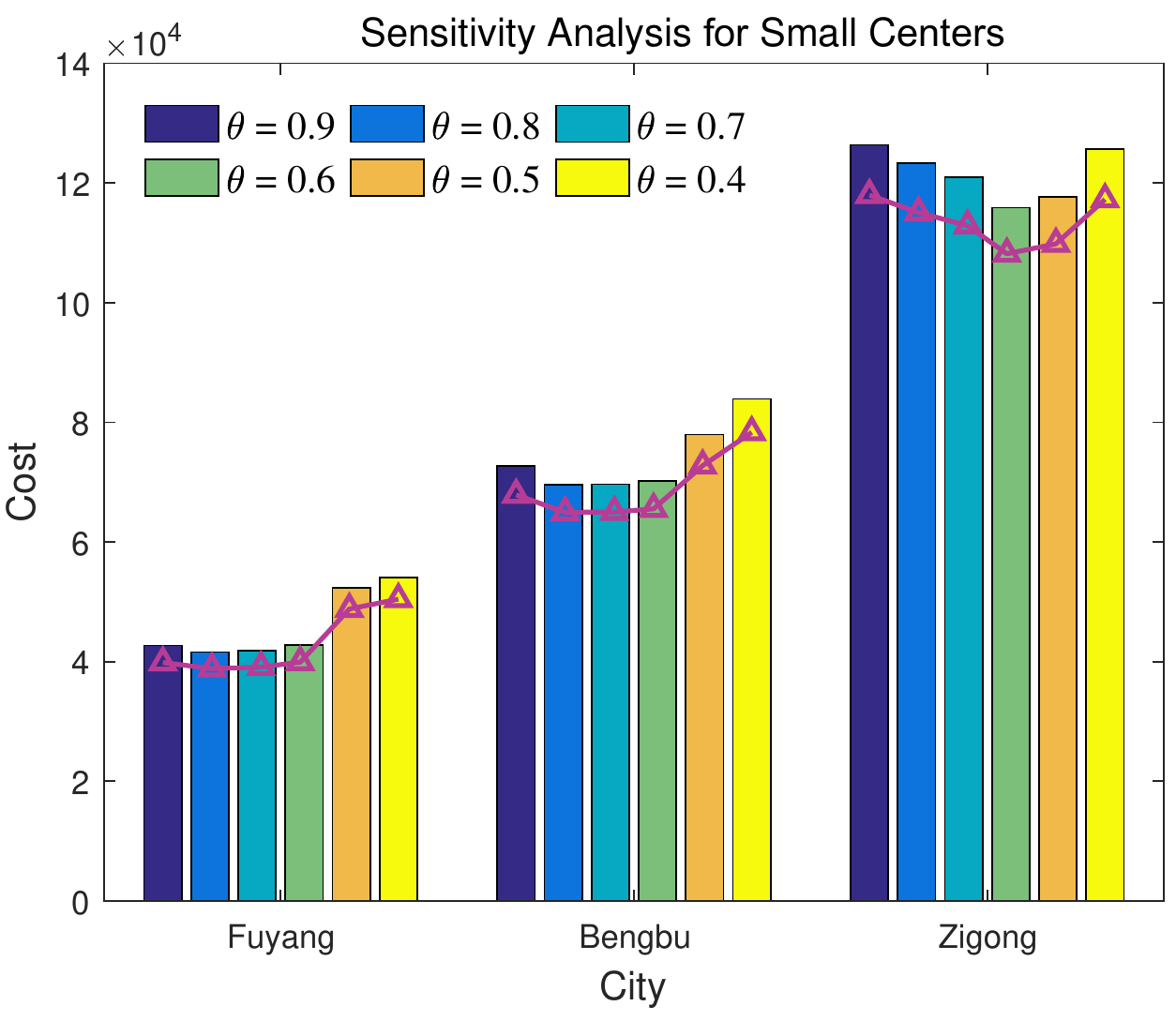} 
	}
	\subfigure{ \label{fig:mid}
		\includegraphics[width=0.30\linewidth]{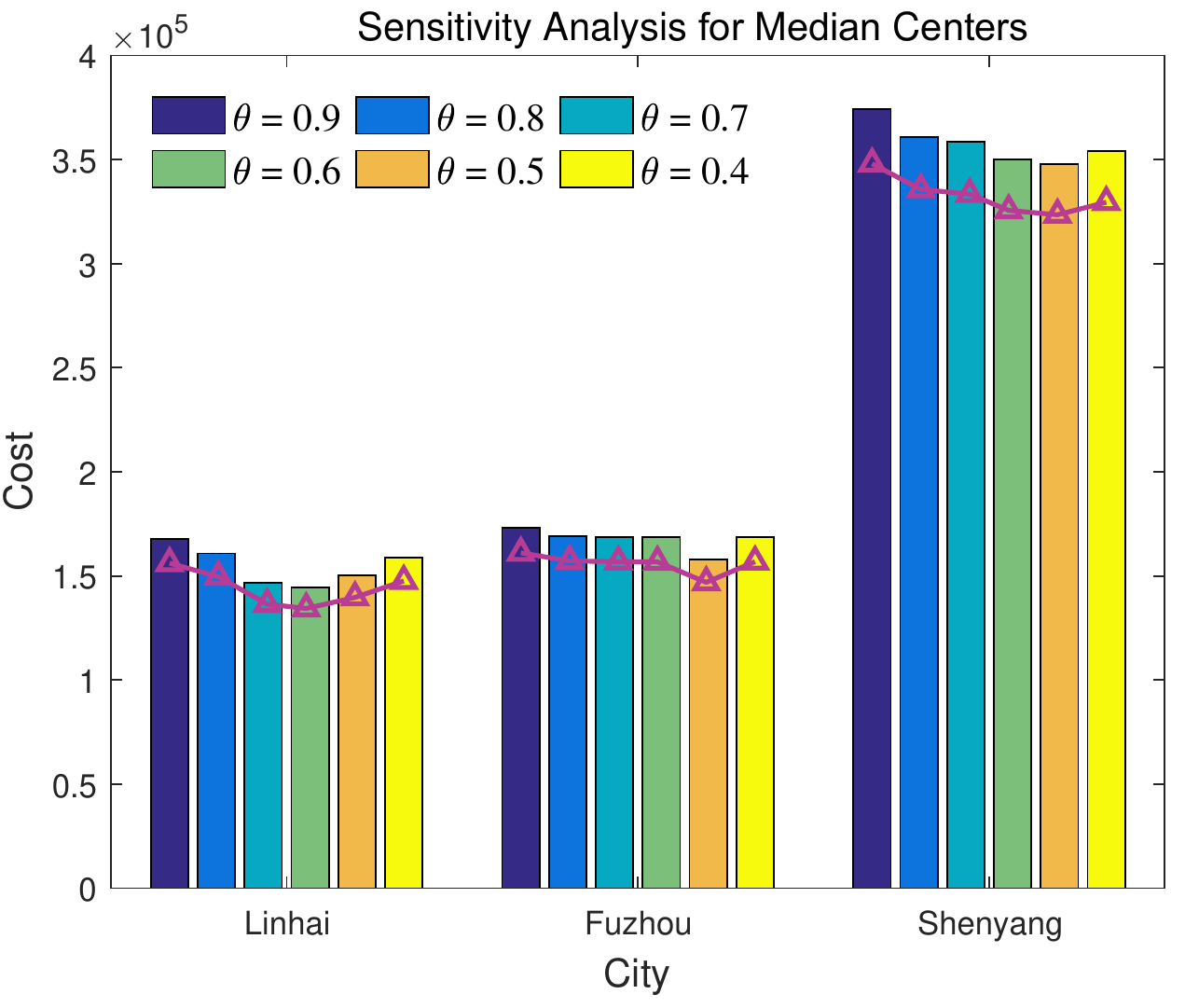} %
	}
	\subfigure{ \label{fig:large}
	\includegraphics[width=0.30\linewidth]{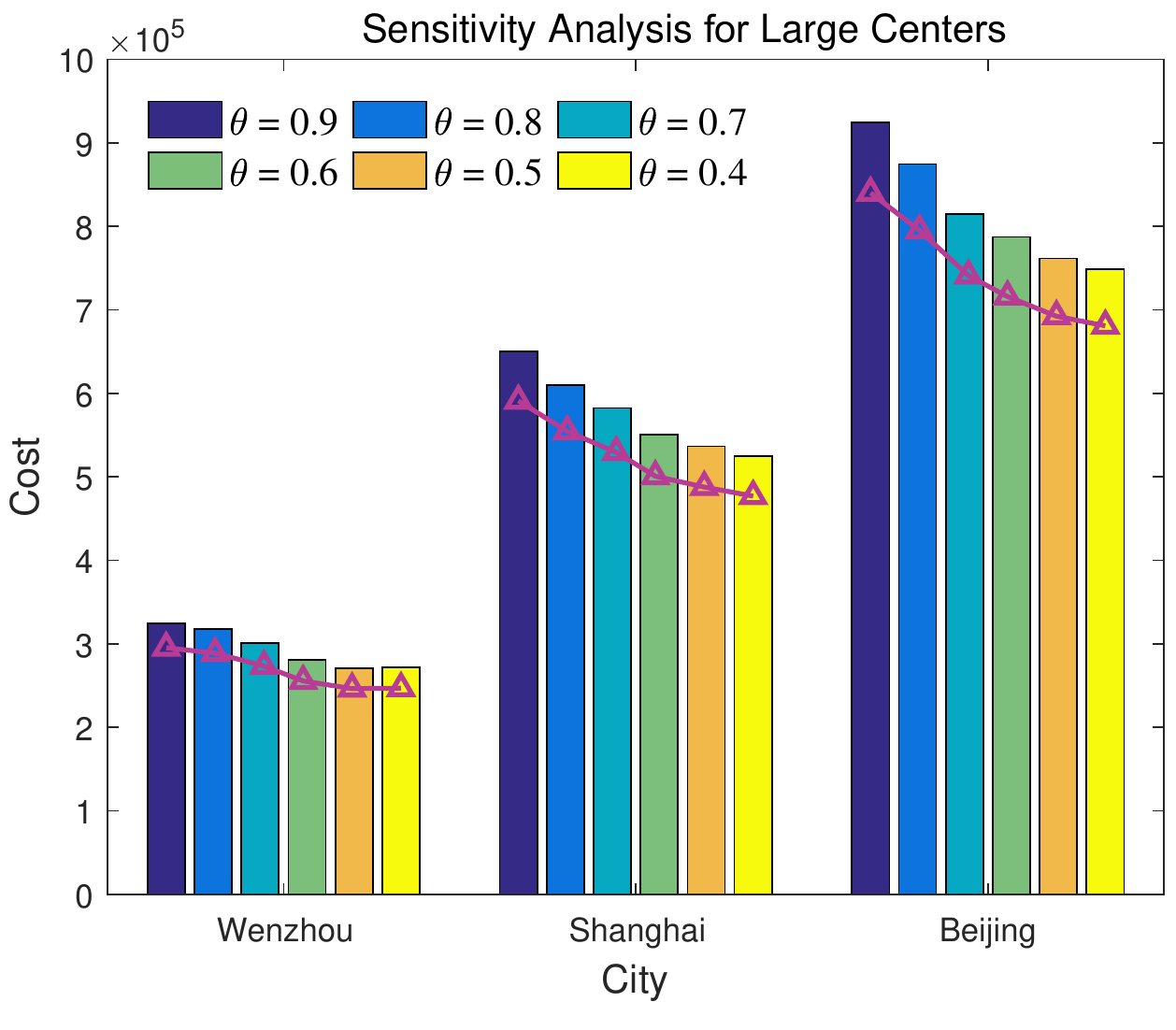} %
	}
	
	\caption{The sensitivity analysis for cities in different sizes}
	\label{fig:sen}
\end{figure}	

Results for different $\theta$ are presented in Figure \ref{fig:sen}. It is observed that for centers of small and middle scale, the cost first decreases and then increases as the value of $\theta$ decreases. For large centers, the cost decreases monotonically as $\theta$ decreases.
In particular, we observe that the optimal threshold increases as the scale of origins increases, which is consistent with the observation in Section \ref{sec:cost_com}, i.e., the larger the origin, the higher percentage of packages shipped by the origin itself. As a result, the threshold of our two-layer LPS-TCN model should be selected based on the size of the origin center for better cost saving.

\section{Conclusion} \label{conclusion}
We propose a novel two-layer LPS-TCN model that emphasizes cost saving. We formulate the LPS-TCN as an MIP and prove its strong NP-hardness. To solve this model, we develop a CG-based algorithm and further enhance it by some problem-specific cutting planes and variable bound tightening techniques. Experiments on realistic instances demonstrate that the LPS-TCN can yield solutions with significant economic benefits. Moreover, the proposed algorithm  significantly outperforms the CPLEX solver in terms of the computation time.

In the future, highly effective algorithms can be studied to tackle large-scale instances of the proposed LPS-TCN model. Moreover, as there exists prediction error in the predicted next-day demand, a model considering uncertain next-day demand can be studied to propose a more reliable transportation plan. In particular, distributionally robust optimization techniques can be applied to address the uncertainty, which exhibits good out-of-sample performance and has been applied to different fields, such as portfolio optimization, appointment scheduling problems, and shortest path problems.

\clearpage
\bibliographystyle{informs2014trsc} 
\bibliography{mybib}

\clearpage
\section*{Online Appendix: Proofs}

\addcontentsline{toc}{section}{Appendices}

\renewcommand{\thesubsection}{\Alph{subsection}}

\subsection{Proof of Theorem \ref{dp_np_hard}}
\label{app_proof_theo1}
\begin{proof}{Proof}
	The problem defined in Theorem \ref{dp_np_hard} is the decision version of the proposed LPS-TCN problem, we denote it as D(LPS-TCN). The proof is based on the reduction from the well-known NP-complete $K$-PARTITION problem to our LPS-TCN problem. 
	
	We show that any instance $I$ of $K$-PARTITION Problem can be polynomially transformed into an equivalent instance $I^\prime$ of D(LPS-TCN) by setting $n = K$. The instance $I^\prime$ of our problem  has an origin center $o$ and a set of destination nodes $\mathcal{V}^1 = \{1,\dots,Km\}$ in the first layer. The demand of each destination $i \in \mathcal{V}^1$ is equal to $\mu_i$. The arc set is  $\mathcal{A}^1 = \{(i,j)~|~i \in \mathcal{V}^1\cup \{o\}, j \in \mathcal{V}^1\setminus \{i\}\}$. There is only one vehicle type with capacity $B$, unit cost per kilometer traveled, and no destination center in the second-layer, i.e., $|\mathcal{K}|=1, \mathcal{V}^2 = \emptyset$. Lastly, we assume that all destinations are evenly distributed on a circular arc centered at the origin center with a radius $R$ and a radian. The radian is small enough such that the chord between any destinations are bounded by a given constant $U_b = {R}/{Km}$. Then, the cost of vehicles on arc $(i,j) \in \mathcal{A}^1$ equals $R$ if $i = o$ and others equals $l_{ij}\leq U_b$. Finally, we set the total cost $C$ to $mR + (K-1)mU_b$. 
	
	We first prove that  the given cost $C$ bounds the number of vehicles used to ship packages to be no more than $m$. Actually, the number of vehicles in any ``YES" instance of D(LPS-TCN) must be $m$, each of which is fully loaded. It can be proved by contradiction as follows. 
	
	If the number of vehicles is smaller than $m$, the capacity is less than $mB$, which is insufficient because the total demand is $\sum_{i=1}^{Km} \mu_i = mB$. If it is larger than $m$, then the cost $C^\prime \ge (m+1)R$. However, 
	$$C = mR + (K-1)mU_b  < mR + (K-1)m*\frac{R}{Km} < (m+1)R \le C^\prime,$$
	which completes the proof. The transportation cost $C^{\prime\prime}$ of a transportation plan using $m$ vehicles is
	$$C^{\prime\prime} = \sum_{p = 1}^m l_p \le mR + m\cdot(K-1) U_b = C,$$
	where $l_p$ is the length of path $p$ used by vehicle $p$. The inequality  holds when the number of arcs of each path is no larger than $K$ and the length of each arc is bounded by $U_b$.

	Next, we prove that in any ``YES'' instance of D(LPS-TCN), all demands must be shipped  without split, i.e., the demand of each destination is shipped completely by one vehicle. Again, this can be proved by contradiction. Without loss of generality, assume there exists a destination $d$  whose demand is shipped by two vehicles, denoted by $k$ and $j$ and all the other demands are shipped without split. For the other $m-2$ vehicles, since $B/(K+1) < \mu_i < B/(K-1)$, the number of destinations visited by each vehicle is exactly $K$. Thus, the total number of destinations visited by $k$-th and $j$-th vehicles is $1 + Km - K(m-2)=2K + 1$, as the node $d$ is visited twice. Hence, there must be a vehicle, either $k$ or $j$, that visits more than $K$ nodes, which is infeasible in our problem due to the limit on the length of path. Consequently, the constructed instance of D(LPS-TCN), $I^\prime$, has to find a partition of destinations denoted by $\mathcal{V}^1_1, \dots, \mathcal{V}^1_m$ such that $\sum_{i \in \mathcal{V}^1_j}\mu_i = B$, which is exactly the decision problems for $K$-PARTITION problem. Hence, if we can find a feasible transportation plan for the LPS-TCN problem, then the instance of $K$-PARTITION problem is a YES-instance, otherwise, the instance of $K$-PARTITION problem is a NO-instance.
	
\end{proof}

\subsection{Proof of Theorem \ref{valid_cut}}
\label{proof:valid_cut}
\begin{proof} {Proof}
	The amount of packages  transported to the destination center $i$ is no smaller than $d_i$. Actually, it equals $d_i$ if no packages for the second layer have been shipped  to $i$ when partial outsourcing is allowed.
	Thus, the smallest number of the vehicles for destination $i$ is $\left\lceil \frac{d_i}{q^*} \right\rceil$, where only the vehicle type with the maximal capacity is used.  
\end{proof}

\subsection{Proof of Theorem \ref{var_red}}
\label{proof:var_red}
\begin{proof} {Proof}
	
	For the ease of exposition, we omit index $k$, i.e. the type of vehicles in this proof. Let path $p = (o,i_1,\dots,i_{n_p})$ be an elementary  path with $n_p \ge 2$, that starts from the origin, i.e., $o$, and visits a set of destination centers $\{i_1,\dots,i_{n_p}\}$. We first show that for any path $p$ with $n_p \ge 2$ and $\bar y_p > 0$ in an optimal solution, the amount of packages delivered to each node in $\{i_1,\dots,i_{n_p}\}$ by  path $p$, denoted by $\{\bar{z}_p^{i_j}\}_{j=1}^{n_p}$, is no more than $q$. It can be proved by contradiction as follows.
	
	Note that the amount of packages delivered to any node on path $p$ is strictly positive. Otherwise, if there exists a node $i_l \in \{i_1,\dots,i_{n_p}\}$ with $\bar{z}_p^{i_l} = 0$, then the path can be shortened to $p^\prime = (o,i_1,\cdots,i_{l-1},i_{l+1},\dots,i_{n_p})$ which has a smaller cost than path $p$. As a result, the path $p$ which contains a node with $\bar{z}_p^{i_j} = 0$  is not in an optimal solution in this case. 
	
	If there exists a node $i_l \in \{i_1,\dots,i_{n_p}\}$ with $\bar{z}_p^{i_l} > q$, we have $\bar{y}_p > 1$, otherwise the capacity of this path is no more than $q$, which is insufficient because the total amount of package is $\sum_{j=1}^{n_p}\bar{z}_p^{i_j} > q$, the inequality holds as $\bar{z}_p^{i_j} > 0, ~\forall i_j \in \{i_1,\dots,i_{n_p}\}$.  Then we can derive another solution with a smaller cost. This solution contains two paths, i.e., $p$ given before and $p^{\prime\prime}= (o,i_l)$. The notations $\hat{y}_p$ and  $\bar{y}_{p^{\prime\prime}}$ count the number of vehicles on path $p$ and $p^{\prime\prime}$ respectively. The scalars $\{\hat{z}_p^{i_j}\}_{j=1}^{n_p}$ and $\{\bar{z}_{p^{\prime\prime}}^{i_l}\}$ denotes the amount of packages shipped to node on different paths.
	
	Let $\bar{y}_{p^{\prime\prime}} = \left\lfloor {\bar{z}_p^{i_l}}/{q} \right\rfloor$ and $\hat{y}_p = \bar{y}_p - \bar{y}_{p^{\prime\prime}}$. Moreover, set $\bar{z}_{p^{\prime\prime}}^{i_l} = \left \lfloor {\bar{z}_p^{i_l}}/{q}\right\rfloor q$, $\hat{z}_p^{i_l} = \bar{z}_p^{i_l} - \left \lfloor {\bar{z}_p^{i_l}}/{q}\right\rfloor q$ and $\hat{z}_p^{i_j} = \bar{z}_p^{i_j}$ for any node $i_j \in \{i_1,\dots,i_{n_p}\}/i_l$.
	
	First we prove that the solution is feasible, i.e., it satisfies that $\sum_{j=1}^{n_p}\hat{z}_p^{i_j} \le \hat{y}_p\cdot q$ and $\bar{z}_{p^{\prime\prime}}^{i_l} \le \bar{y}_{p^{\prime\prime}} \cdot q$.
	The second inequality is trivial due to the definition. We next prove the first inequality. We have that  
	
	$$\sum_{j=1}^{n_p}\hat{z}_p^{i_j} = \sum_{j=1}^{n_p}\bar{z}_p^{i_j} - \left \lfloor \frac{\bar{z}_p^{i_l}}{q}\right \rfloor \cdot q \le \bar{y}_p\cdot q - \left\lfloor \frac{\bar{z}_p^{i_l}}{q}\right\rfloor \cdot q = (\bar{y}_p-\bar{y}_{p^{\prime\prime}})q = \hat{y}_p \cdot q$$
	where the first inequality holds due to the fact $\sum_{j=1}^{n_p}\bar{z}_p^{i_j} \le \bar{y}_p\cdot q.$ Hence, the constructed solution is feasible.

	Then we prove that the solution has a less cost than that of the primal solution.	Let $c_{new}$ be the cost of this new solution, we have
	\begin{equation*}
		\begin{aligned}
			c_{new} = l_{oi_l}\cdot \bar{y}_{p^{\prime\prime}}\cdot c + l_p \cdot \hat{y}_p\cdot c = l_{oi_l}\cdot \bar{y}_{p^{\prime\prime}}\cdot c + l_p \cdot (\bar{y}_p - \bar{y}_{p^{\prime\prime}})\cdot c = l_p \cdot \bar{y}_p \cdot c - (l_p - l_{oi_l})\cdot \bar{y}_{p^{\prime\prime}}\cdot c < l_p \cdot \bar{y}_p \cdot c,
		\end{aligned}
	\end{equation*}
	where the last inequality holds due to Assumption \ref{assum:org} and \ref{assum-tra}, i.e. 
	\begin{equation}
		\begin{aligned}
			\label{triequality}
			l_{oi_l} < l_{oi_{l-1}}+l_{i_{l-1}i_l} < l_{{oi_{l-2}}}+  l_{i_{l-2}i_{l-1}}+l_{i_{l-1}i_l} < \cdots < l_{oi_1} + l_{i_1i_2} + \cdots + l_{i_{l-1}i_l} \le l_p.
		\end{aligned}
	\end{equation}
	Then we complete the proof.

	Next we prove another part of Theorem \ref{var_red} by induction on $n_p$, i.e. the value of integer variables $\bar{y}_p \le 1$ for path with $n_p \ge 2$  in an optimal solution.

	\textbf{Step 1:} 
	
	We first show that the statements holds for the smallest number $n_p = 2$. Let $p =(o,i_1,i_2)$ be the path.
	
	Note that we always have $\bar{z}^{i_1}_p < q$ and $\bar{z}^{i_2}_p < q$. Then we prove Theorem \ref{var_red} by contradiction. Assume that $\bar{y}_p > 1$ in an optimal solution. As $\bar{z}^{i_1}_p < q$ and $\bar{z}^{i_2}_p < q$, we have $\bar{z}^{i_1}_p+\bar{z}^{i_2}_p < 2q$, consequently, $\bar{y}_p = 2$. Without loss of generality, assume that $\bar{z}^{i_1}_p < \bar{z}^{i_2}_p$. Then we can use one vehicle to transport packages to ${i_1}$ with number being $\bar{z}^{i_1}_p$ and ${i_2}$ with  $q - \bar{z}^{i_1}_p$ by path $p$, and use another one vehicle to transport  packages to $i_2$ with number being $ \bar{z}^{i_2}_p + \bar{z}^{i_1}_p - q$ by path $p^\prime = (o,i_2)$. As a result, we have
	\begin{equation*}
		\begin{aligned}
			c_{new} &= l_{p^\prime}\cdot c + l_p \cdot c < 2 \cdot l_p\cdot c,\\
		\end{aligned}
	\end{equation*}
	where the last inequality holds due to Assumption \ref{assum-tra}. Hence, $y_p > 1$ is not in an optimal solution and we derive a contradiction. 
	
	\textbf{Step 2:} Assume that the statement holds for arbitrary natural number $n_p = n$.
	
	\textbf{Step 3:} In this step we prove that the statement holds for $n_p = n + 1$, i.e., the path $p = (o,i_1,\cdots,i_n,i_{n+1})$.
	
	
	Assume that $\bar{y}_p > 1$ in an optimal solution. As $\bar{z}^{i_{n+1}}_p < q$, if $\sum_{j=1}^{n}\bar{z}^{i_j}_p < q$, we have $\sum_{j=1}^{n}\bar{z}^{i_j}_p +\bar{z}^{i_{n+1}}_p < 2q$, consequently, $\bar{y}_p = 2$. Without loss of generality, assume that $\sum_{j=1}^{n}\bar{z}^{i_j}_p < \bar{z}^{i_{n+1}}_p$. We prove that we can derive another solution with a less cost. This solution contains two paths, i.e., $p$ given before and $p^{\prime}= (o,i_{n+1})$. The $\hat{y}_p$ and  $\bar{y}_{p^{\prime}}$ count the number of vehicles on path $p$ and $p^{\prime}$ respectively. The  $\{\hat{z}_p^{i_j}\}_{j=1}^{n+1}$ and $\{\bar{z}_{p^{\prime}}^{i_{n+1}}\}$ denote the amount of packages shipped to node on different paths.
	
	Let $\bar{y}_{p^{\prime}} = 1 $ and $\hat{y}_p = 1$. Moreover, set
	$\hat{z}_p^{i_j} = \bar{z}_p^{i_j}$ for  $i_j \in \{i_1,\dots,i_n\}$,
	$\hat{z}_p^{i_{n+1}} = q - \sum_{j=1}^n \bar{z}_p^{i_j}$ and 
	$\bar{z}_{p^{\prime}}^{i_{n+1}} = \bar{z}_p^{i_{n+1}} - \hat{z}_p^{i_{n+1}}$. Obviously, this solution is feasible, that is, it satisfies that $\sum_{j=1}^{n+1}\hat{z}_p^{i_j} \le q$ and $\bar{z}_{p^{\prime}}^{i_{n+1}} = \bar{z}_p^{i_{n+1}} - \hat{z}_p^{i_{n+1}} \le  q$.
	
	Let $c_{new}$ denote the cost of the constructed solution, it follows that
	\begin{equation*}
		\begin{aligned}
			c_{new} &= l_{p^\prime}\cdot c + l_p\cdot c < 2l_p\cdot c,\\
		\end{aligned}
	\end{equation*}
	where the last inequality holds due to Assumption \ref{assum:org} and \ref{assum-tra}, see \eqref{triequality} for details. Hence, $y_p > 1$ is not in an optimal solution in this case. 
	
	If $\sum_{j=1}^n\bar{z}^{i_j}_p \ge q$, we can also find a feasible solution with a less cost than current solution. It contains the given path $p$ and a different path $p^{\prime\prime} = (o,i_1,\dots,i_n)$. 
	Similar to the proof in the first case, we have decision variables $\bar{y}_{p^{\prime\prime}}$, $\hat{y}_p$, $\{\bar{z}^{i_j}_{p^{\prime\prime}}\}_{j=1}^n$ and $\{\hat{z}^{i_j}_{p}\}_{j=1}^{n+1}$.
	
	Let $\bar{y}_{p^{\prime\prime}} = \left\lfloor {\sum_{j=1}^n\bar{z}^{i_j}_p}/{q} \right\rfloor$ and $\hat{y}_p = \bar{y}_p - \bar{y}_{p^{\prime\prime}}$. Moreover, we obtain a group of $\{\bar{z}^{i_j}_{p^{\prime\prime}}\}_{j=1}^n$ such that $\sum_{j=1}^n\bar{z}^{i_j}_{p^{\prime\prime}} = q\left\lfloor {\sum_{j=1}^n\bar{z}^{i_j}_p}/{q} \right\rfloor$ by adjusting the amount of packages to node $i_j \in \{i_1,\dots,i_n\}$ on path $p$. Note that, this group follows that $\bar{z}^{i_j}_{p^{\prime\prime}} \le \bar{z}^{i_j}_p, ~\forall~ {i_j} \in \{i_1,\dots,i_n\}$. Then we set $\hat{z}_p^{i_j} = \bar{z}_p^{i_j}-  \bar{z}^{i_j}_{p_{\prime\prime}}, ~\forall~ i_j \in \{i_1,\dots,i_n\}$ and $\hat{z}_p^{i_{n+1}} = \bar{z}_p^{i_{n+1}}$. 
	
	We first prove that the solution satisfies that $\sum_{j=1}^{n+1}\hat{z}_p^{i_j} \le \hat{y}_p\cdot q$ and $\sum_{j=1}^n \bar{z}_{p^{\prime\prime}}^{i_j}  \le \bar{y}_{p^{\prime\prime}} \cdot q$.
	The second inequality is trivial due to the definition. We next prove the first inequality. We have that  
	
	$$\sum_{j=1}^{n+1}\hat{z}_p^{i_j} = \sum_{j=1}^{n+1}\bar{z}_p^{i_j} - \sum_{j=1}^{n}\bar{z}_{p^{\prime\prime}}^{i_j} = \sum_{j=1}^{n+1}\bar{z}_p^{i_j} - \left\lfloor \frac{\sum_{j=1}^{n}\bar{z}^{i_j}_{p}}{q}\right\rfloor \cdot q \le \bar{y}_p\cdot q -\bar{y}_p^{\prime\prime} \cdot q = (\bar{y}_p-\bar{y}_p^{\prime\prime})q = \hat{y}_p \cdot q.$$
	Hence, the constructed solution is feasible. 
	
	Let $c_{new}$ be the cost of the constructed solution, we have 
	\begin{equation*}
		\begin{aligned}
			c_{new} = l_{p^{\prime\prime}}\cdot \bar{y}_{p^{\prime\prime}}\cdot c + l_p \cdot \hat{y}_p\cdot  c
			=l_{p^{\prime\prime}} \cdot \bar{y}_{p^{\prime\prime}}\cdot c + l_p(\bar{y}_p -\bar{y}_{p^{\prime\prime}})c 
			= l_p\cdot \bar{y}_p\cdot c - (l_p - l_{p^{\prime\prime}})\bar{y}_{p^{\prime\prime}} c < l_p\cdot \bar{y}_p\cdot c.
		\end{aligned}
	\end{equation*}
	Hence, we derive a contradiction and $y_p > 1$ is also not in an optimal solution in this case.

\end{proof}

\subsection{Proof of Theorem \ref{bound:one_arc}}
\label{proof:bound:one_arc}
\begin{proof}{Proof}
	For any optimal solution $(\bar x, \bar y)$ to \eqref{pri-model} and $u_k^*$ to \eqref{add_bound}, if $\bar{y}_p^k > u_k^*$ for any path $p$ with only one arc, according to problem \eqref{add_bound}, we can reduce $\bar{y}_p^k$ to $\bar{y}_p^k - u_k^*$ by finding a group of vehicles  to replace $u_k^*$ of type $k$ vehicles on path $p$. Moreover, the total cost of this group of vehicles is no larger than $c_ku_k^*$ but the total capacity is no less than $q_ku_k^*$, which holds due to the constraints \eqref{con:cost} and \eqref{con:cap}. As a result, this replacement does not increase the cost. Note that the capacity of each vehicle in the group should be larger than $q_k$, otherwise, it may lead to a loop of replacements, i.e., if we use vehicles with a smaller capacity, i.e., type $i < k$ to replace the type $k$ vehicle, it may make the number of type $i$ vehicles larger than its upper bound $u_i^*$, then we have to find another group of vehicles to replace type $i$ vehicles, which may bring the number of vehicle type $k$ back to its original value. 
\end{proof}


\end{document}